\documentclass[showpacs,preprintnumbers,amsmath,amssymb,floatfix]{revtex4}

\usepackage{color}   
\usepackage{graphicx}
\usepackage{dcolumn}
\usepackage{bm}

\begin{document}

\def\bbox#1{\hbox{\boldmath${#1}$}}
\def\blambda{{\hbox{\boldmath $\lambda$}}}
\def\eeta{{\hbox{\boldmath $\eta$}}}
\def\bxi{{\hbox{\boldmath $\xi$}}}
\def\bzeta{{\hbox{\boldmath $\zeta$}}}

\title{ The Momentum Kick Model Description of the Near-Side Ridge and
Jet Quenching}

\author{Cheuk-Yin Wong}

\affiliation{Physics Division, Oak Ridge National Laboratory, 
Oak Ridge, TN\footnote{wongc@ornl.gov} 37831}

\date{\today}

\begin{abstract}

In the momentum kick model, a near-side jet emerges near the surface,
kicks medium partons, loses energy, and fragments into the trigger
particle and fragmentation products.  The kicked medium partons
subsequently materialize as the observed ridge particles, which carry
direct information on the magnitude of the momentum kick and the
initial parton momentum distribution at the moment of jet-(medium
parton) collisions.  The initial parton momentum distribution
extracted from the STAR ridge data for central AuAu collisions at
$\sqrt{s_{NN}}=200$ GeV has a thermal-like transverse momentum
distribution and a rapidity plateau structure with a relatively flat
distribution at mid-rapidity and sharp kinematic boundaries at large
rapidities.  Such a rapidity plateau structure may arise from particle
production in flux tubes, as color charges and anti-color charges
separate at high energies.  The centrality dependence of the ridge
yield and the degree of jet quenching can be consistently described by
the momentum kick model.

\end{abstract}


\maketitle


\section{Introduction}

In central high-energy heavy-ion collisions, jets are produced in
nucleon-nucleon collisions and they interact with the dense medium
produced in the interacting region.  Depending on the relative
azimuthal angle relative to the trigger particle, observed high-$p_t$
jets can be classified as near-side jets or away-side jets.  An
away-side jet is associated with a broad cone of particles pointing
azimuthally opposite to the trigger particle direction.  The strong
attenuation of the away-side jet in its passage through the produced
dense matter is one of the many notable experimental observations in
relativistic heavy-ion collisions and is a signature for the
production of the strongly-coupled quark-gluon plasma
\cite{Adc04,Ada05a,Ars04,Bac04}.

On the other hand, a near-side jet is characterized by the presence of
associated particles within a narrow azimuthal angle along the trigger
particle direction.  It retains many of the characteristics of the
associated fragmentation products as those of a jet in $pp$ and peripheral
heavy-ion collisions.  The near-side jet occurs when the high-$p_t$
jet emerges near the surface of the produced parton medium.

Recently, the STAR Collaboration
\cite{Ada05,Ada06,Put07,Bie07,Wan07,Bie07a,Abe07,Mol07,Lon07,Nat08,
  Net08} observed a $\Delta \phi$-$\Delta \eta$ correlation of
particles associated with a near-side, high-$p_t$ trigger particle in
central AuAu collisions at $\sqrt{s_{NN}}=200$ GeV at RHIC, where
$\Delta \phi$ and $\Delta \eta$ are the azimuthal angle and
pseudorapidity differences measured relative to the trigger particle,
respectively. Particles associated with the near-side jet can be
decomposed into a ``jet component'', corresponding to fragmentation
products of the near-side jet at $(\Delta \phi, \Delta \eta)$$
\sim$(0,0), and a ``ridge component'' at $\Delta\phi$$ \sim$0 with a
ridge structure in $\Delta \eta$.  Similar $\Delta \phi$-$\Delta \eta$
correlations associated with a near-side jet have also been observed
by the PHENIX Collaboration \cite{Ada08,Mcc08} and the PHOBOS
Collaboration \cite{Wen08}.  While many theoretical models
\cite{Won07,Won07a,Won08,Hwa03,Chi05,Hwa07,Rom07,Vol05,Arm04,Shu07,Pan07,Dum07,Dum08,Miz08,Jia08,Gav08,Gav08a}
have been proposed to discuss the jet structure and related phenomena,
the ridge phenomenon has not yet been fully understood.

Previously, a momentum kick model was put forth to explain the ridge
phenomenon \cite{Won07,Won07a,Won08}.  The model assumes that a
near-side jet occurs near the surface, kicks medium partons, loses
energy along its way, and fragments into the trigger and its
associated fragmentation products (the ``jet component'') (Fig.\ 1).
Those medium partons that are kicked by the jet acquire a momentum
kick along the jet direction.  They subsequently materialize by
parton-hadron duality as ridge particles in the ``ridge component''
(Fig. 1).  They carry direct information on the momentum distribution
of the medium partons at the moment of jet-(medium parton) collisions,
for which not much information has been obtained from direct
experimental measurements.  As the early state of the medium partons
is an important physical quantity, it is therefore useful to examine
the early parton momentum distribution using the momentum kick model.

A previous momentum kick model analysis gave theoretical results in
qualitative agreement with experimental data \cite{Won07}.  We arrived
at the interesting observation that at the moment of jet-(medium parton)
collisions the parton transverse slope parameter $T$ is slightly
higher and the rapidity width substantially greater than corresponding
quantities of their evolution products at the end-point of the
nucleus-nucleus collision.  We would like to refine the model and give
a quantitative comparison with experiment.  We also wish to explore
the early parton momentum distribution over a wider kinematic range of
different transverse momenta and rapidities to search for interesting
and novel features of the initial parton momentum distribution.

\begin{figure} [h]
\includegraphics[angle=0,scale=0.50]{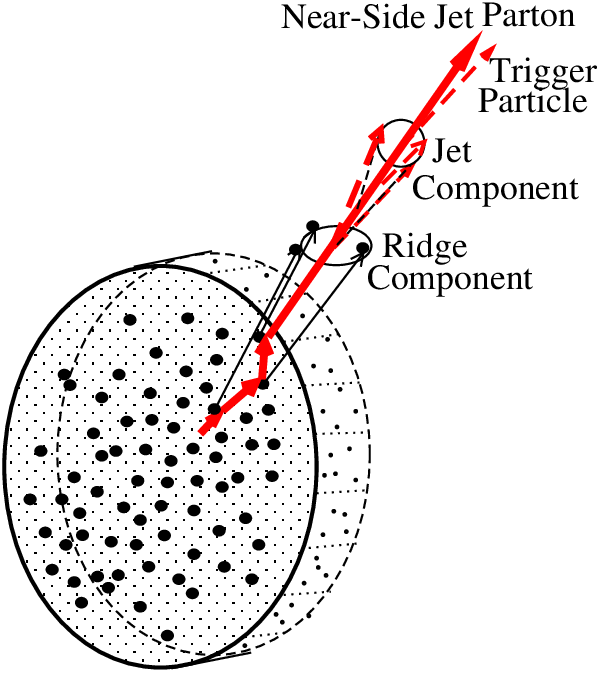}
\vspace*{0.0cm} 
\caption{ (Color online) Schematic representation of the momentum kick
model.  A near-side jet parton (represented by adjoining thick arrows)
occurs in a dense medium, whose partons are represented by solid
circular points.  The jet parton kicks many medium partons, loses
energy, radiates, and fragments into the trigger particle and
associated ``jet component'' particles.  The medium partons that are
kicked by the jet parton acquire a momentum kick along the jet
direction and materialize as associated ``ridge component''
particles. Not shown in the figure is the away-side jet opposite to
the near-side jet. }
\end{figure}

We shall show that the extracted early parton distribution has a
plateau rapidity structure.  Rapidity distributions in the form of a
plateau have been known in QCD particle production processes both
experimentally and theoretically.  Experimental evidence for a plateau
rapidity distributions along the sphericity axis or the thrust axis
has been found in $\pi^\pm$ production in $e^+$-$e^-$ annihilation
\cite{Aih88,Hof88,Pet88,Abe99,Abr99,Not1}.  The rapidity distributions
for $K^\pm$ production also show a plateau structure with a depression
in an extended region around $y\sim 0$.  As a function of energy, the
shape of the rapidity distribution for the sum of all particles
produced in $e^+$-$e^-$ annihilation is a plateau with either a flat
distribution or a small depression at $y\sim 0$ \cite{Abr99}.  The
width of the rapidity plateau increases as a function of increasing
center-of mass energy.

Theoretically, the rapidity plateau structure has been known in many
earlier investigations of particle production processes in QCD, when a
quark pulls away from an anti-quark at high energies
\cite{Cas74,Bjo83,Won91,Won94,And83}.  The theoretical basis in the
work of Refs.\ \cite{Cas74,Bjo83,Won91,Won94} comes from the
approximate connection between QCD and QED2
\cite{Sch62,Low71,Col75,Col76}.  We would like to show more explicitly
here how the transverse confinement in a flux tube allows one to
establish an approximate connection between the field theories of QED2
and QCD in high energy processes.  Using such a connection, we
wish to review here how the rapidity plateau structure occurs when a
color charge separates from an anti-color charge at high energies.

The early stage of the nucleus-nucleus collision comprise of many
simultaneous elementary particle production processes involving a
quark pulling away from an anti-quark (or $qq$ diquark) at high
energies.  As elementary processes lead to plateau rapidity
distributions, the rapidity distribution of the medium partons at the
early stage of the nucleus-nucleus collision should retain the
rapidity plateau characteristics.

The parton momentum distribution is only one aspect of the momentum
kick model.  The momentum loss of the jet in its passage through the
medium is another important aspect.  While many theoretical treatments
of the jet quenching phenomenon have been presented previously
\cite{Jetxxx}, the investigation of the jet quenching phenomenon in
connection with ridge particles associated with the jet will provide a
different and complimentary perspective.  The ridge yield and the
quenching of the emerging jet will depend on the number of medium
partons kicked by the jet along its way.  A successful simultaneous
description of the centrality dependence of the ridge yield and jet
quenching will provide a consistent picture of the interaction between
a jet and the medium.  It will also pave the way for a future Monte
Carlo implementation of the momentum kick model where many refinements
and improvements can be included.

In this paper, we shall limit our attention to particles associated
with the near-side jet.  In the context of the present investigation,
the ridge particle momentum distributions in nucleus-nucleus
collisions refer implicitly to those measured on a ``per jet trigger''
basis, unless indicated otherwise.  For brevity of terminology, the
term ``jet'' will be used both for the parent ``jet parton'' that
passes through the medium and also for the daughter ``jet component''
of associated particles.  The ambiguity of the meaning of the term can
be easily resolved by context.

This paper is organized as follows.  In Section II we summarize the
momentum kick model and relate the ridge yield to the number of kicked
medium partons.  In Section III, we give the relationship between the
initial and final parton momentum distributions under the action of a
momentum kick.  In Section IV, we specify how the initial parton
momentum distribution is parametrized.  In order to determine the
initial parton momentum distribution from the observed total particle
distribution in central AuAu collisions, we parametrize in Section V
the jet momentum distribution associated with the near-side jet in
$pp$ collisions.  The momentum distribution of mid-rapidity associated
particles in central AuAu collisions at $\sqrt{s_{NN}}=200$ GeV are
then analyzed in Section VI.\ In Section\ VII, we display explicitly
the initial parton momentum distribution extracted from the ridge
data.  In Section VIII, we examine the connection between QCD and QED2
in the presence of transverse confinement and study the origin of the
rapidity plateau when a color charge separates from an anti-color
charge at high energies.  In Section IX, the field theory of bosonized
QED2 is then used to study particle production as an initial-value
problem.  The evolution of the medium parton momentum distribution is
discussed in Section X.  In Section XI, we turn our attention to the
propagation of the jet and the dependence of the jet fragmentation
function on the number of jet-(medium parton) collisions.  In Section
XII, the centrality dependence the ridge yield and jet quenching is
examined.  In Section XIII, we examine the dependence of the ridge
yield on the energy and mass number of the colliding nuclei.  In
Section XIV, we present our conclusions and discussions.
 
\section{The Momentum Kick Model}

It should be pointed out on the outset that the interaction between a
jet and the medium can be described by representing the medium either
as fields or as particles.  In our momentum kick model, we choose to
represent the medium as particles.  We describe the interaction
between the jet and the medium in terms of jet-(medium parton)
collisions, from which each collided medium parton receives a momentum
kick and subsequently materializes as a ridge particle.  We have been
guided to such a particle description because of the strong color
screening in a dense color medium \cite{Kac05,Won02}.  The presence of
the azimuthal kinematic correlation between the jet and the ridge
particles lends additional support to the concept of jet-(medium
parton) collisions as a central element of the phenomenon.

As depicted in Fig. 1, the main contents of the momentum kick model
can be briefly summarized as follows:

\begin{enumerate}

\item
A near-side jet parton emerges near the medium surface and the jet
parton collides with medium partons on its way to the detector.  It
loses energy by collisions and gluon radiation. It subsequently
fragments into the trigger particle and other associated fragmentation
products.

\item
Each jet-(medium parton) collision imparts a momentum kick ${\bf q}$
to the initial medium parton of momentum ${\bf p}_i$ in the general
direction of the trigger particle to change it to the final parton
momentum ${\bf p}\equiv {\bf p}_f={\bf p}_i+{\bf q}$, and it modifies
the normalized initial parton momentum distribution $dF/d{\bf p}_i$ to
become the final parton momentum distribution $dF/d{\bf p}_f$.  The
kicked partons subsequently materialize by parton-hadron duality as
ridge particles.

\end{enumerate}

Based on the above picture, we can describe the jet and the kicked
particles in quantitative terms.  We consider a nucleus-nucleus
collision at a given impact parameter $b$ with $N_{\rm bin}$ binary
nucleon-nucleon collisions, and we label a binary collision by the
index $i$.  For the $i$th binary collision, there is a jet parton
distribution $dN_{ j}^i/d{\bf p}_j$ where the subscript $j$ stands for
the ``jet parton''.  The sum over all binary collisions for a given
impact parameter leads to the total jet parton distribution
$dN_j/d{\bf p}_j$ defined by
\begin{eqnarray}
\label{Nj}
\frac{dN_j}{d{\bf p}_j}=\sum_{i=1}^{N_{\rm bin}} 
\frac{dN_j^i}{d{\bf p}_j}.
\end{eqnarray}
In a single $pp$ collision, the yield of a
trigger particle with momentum ${\bf p}_{\rm trig}$ is
\begin{eqnarray}
\frac{dN^{pp}}{d{\bf p}_{\rm trig}}( {\bf p}_{\rm trig})
= \int d {\bf p}_j \frac{dN^{pp}}{d{\bf p}_{j}} 
{\tilde D} ({ \bf p}_{\rm trig}, {\bf p}_j), 
\end{eqnarray}
where ${\tilde D}({\bf p}_{\rm trig}, {\bf p}_j)$ is the fragmentation
function for fragmenting a trigger hadron of momentum ${\bf p}_{\rm
  trig}$ out of a parent jet parton of momentum ${\bf p}_j$. For
convenience of accounting in nucleus-nucleus collisions for a fixed
${\bf p}_{\rm trig}$, we rescale the fragmentation function by
dividing the above equation by the quantity on the lefthand side,
$[{dN^{pp}}/{d{\bf p}_{\rm trig}} ( {\bf p}_{\rm trig})]$, to change
the above equation to
\begin{eqnarray}
1
= \int d {\bf p}_j \frac{dN^{pp}}{d{\bf p}_{j}} 
{D} ({ \bf p}_{\rm trig}, {\bf p}_j), 
\end{eqnarray}
where the re-normalized
fragmentation function $D({\bf p}_{\rm trig}, {\bf p}_j)$ is
\begin{eqnarray}
D( {\bf p}_{\rm trig}, {\bf p}_j)= {\tilde D} ({ \bf p}_{\rm trig},
{\bf p}_j) / [{dN^{pp}}/{d{\bf p}_{\rm trig}}( { \bf p}_{\rm trig})].
\end{eqnarray}
Using the fragmentation function normalized in this manner, a binary
nucleon-nucleon collision (the $i$-th binary collision, say) produces
a single trigger particle at the momentum ${\bf p}_{\rm trig}$, in the
absence of jet-medium interactions,
\begin{eqnarray}
\label{dnorm}
\int d{\bf p}_j \frac{dN_{ j}^i}{d{\bf p}_j} D({\bf p}_{\rm trig};
{\bf p}_j) = ({\rm unquenched~} N_{\rm trig}
{\rm~arising~from~the~}i{\rm th~binary~collision})=1.
\end{eqnarray}
With the total jet source distribution $dN_j/d{\bf p}_j$ coming from
all binary collisions in a nucleus-nucleus collision, we have
\begin{eqnarray}
N_{\rm bin}=
\int d{\bf p}_j \frac{dN_{ j}}{d{\bf p}_j} D({\bf p}_{\rm trig}
{\bf p}_j). 
\end{eqnarray}
In the presence of jet-medium interactions, the total number of
trigger particles $N_{\rm trig}$ with momentum ${\bf p}_{\rm trig}$ is
\begin{eqnarray}
\label{trig}
N_{\rm trig} \!\!= \!\!\int \! d{\bf p}_j \frac {dN_j}{d{\bf p}_j} 
\!\sum_{N\!=\!0}^{N_{\rm max}}
\!P(N) 
e^{-\zeta_a N} 
 D({\bf p}_{\rm trig};{\bf p}_j \!-\! \sum_{n = 1}^N {\bf q}_n 
- {\bf \Delta}_r ),
\end{eqnarray}
where $N$ is the number of medium partons kicked by a jet of momentum
${\bf p}_j$ along its way, $N_{\rm max}$ is the maximum $N$
considered, and $P(N)$ is a probability distribution of $N$,
normalized by $\sum_{N=0}^{N_{\rm max}}P(N)=1$.  The factor
$e^{-\zeta_a N}$ describes the absorption of the jet due to the
inelastic fraction of jet-(medium parton) collisions.  The quantity
${\bf q}_n$ is the momentum kick on the medium parton due to the $n$th
jet-(medium parton) collision, and ${\bf \Delta}_r$ is the jet
momentum loss due to the gluon radiation of the jet.  We shall
postpone our discussion of $P(N)$ to Section XII.  It suffices to
indicate here that $P(N)$ depends on the medium parton density along
the trajectory and the jet-(medium parton) scattering cross section.

As is obvious from Eq.\ (\ref{trig}), the number of trigger particles
$N_{\rm trig}$ in a nucleus-nucleus collision (with the momentum ${\bf
  p}_{\rm trig}$) will be equal to the number of binary collisions
$N_{\rm bin}$ in the absence of any jet-medium interaction,
\begin{eqnarray}
N_{\rm trig}(\{\zeta_a,{\bf q}_n,{\bf \Delta}_r\}=0) = N_{\rm bin}.
\end{eqnarray}
The ratio of $N_{\rm trig}(\{\zeta_a,{\bf q}_n,{\bf \Delta}_r\}\ne 0)$
in a nucleus-nucleus collision with respect to $N_{\rm
trig}(\{\zeta_a,{\bf q}_n,{\bf \Delta}_r\}=0)$ in the absence of any
jet-medium interaction is the $R_{AA}$ measure of jet quenching,
\begin{eqnarray}
R_{AA}=\frac{1}{N_{\rm bin}}\!\!\int \! d{\bf p}_j \frac {dN_j}{d{\bf
p}_j} \!\sum_{N\!=\!0}^{N_{\rm max}} \!P(N) e^{-\zeta_a N} D({\bf
p}_{\rm trig};{\bf p}_j \!-\! \sum_{n = 1}^N {\bf q}_n - {\bf
\Delta}_r ).  
\end{eqnarray}
Because the kicked partons are identified as ridge particles by
parton-hadron duality and two-third of the produced hadrons are
charged, the distribution of associated ridge particles is therefore
\begin{eqnarray}
\label{rid}
\frac{dN_{\rm ridge}^{AA}}{d{\bf p}}
&=&\int d{\bf p}_j 
\frac{dN_j}{d{\bf p}_j} \sum_{N=1}^{N_{\rm max}}
P(N) e^{-\zeta_a N} 
D({\bf p}_{\rm trig};{\bf p}_j - \sum_{n=1}^N {\bf q}_n
- {\bf \Delta}_r )
\left \{ \frac {2}{3} \sum_{n = 1}^N  f_{Rn} \frac{dF_n}{d{\bf p}}({\bf q}_n)
\right \}, 
\end{eqnarray}
where $0<f_{Rn}\le 1$ is the ridge attenuation factor for the $n$-th
kicked parton to reach the detector and $dF_n/d{\bf p}$ is the
normalized momentum distribution of the $n$-th kicked medium parton,
normalized to $\int d{\bf p} dF_n/d{\bf p} =1$.  We note that the
righthand sides of Eq.\ (\ref{trig}) and (\ref{rid}) differ only by
the quantity in the curly bracket.  It is convenient to define the
expectation value $\langle {\cal O}\rangle$ of a quantity ${\cal O}$
in the presence of the jet distribution, jet momentum loss, and jet
fragmentation by
\begin{eqnarray}
\label{expt}
\langle {\cal O}\rangle
&=&
\int d{\bf p}_j 
\frac{dN_j}{d{\bf p}_j} \sum_{N=1}^{N_{\rm max}}
P(N) e^{-\zeta_a N} 
D({\bf p}_{\rm trig};{\bf p}_j - \sum_{n=1}^N {\bf q}_n
- {\bf \Delta}_r )
{\cal O}
\nonumber \\
&\div&
\int d{\bf p}_j 
\frac{dN_j}{d{\bf p}_j} \sum_{N=1}^{N_{\rm max}}
P(N) e^{-\zeta_a N} 
D({\bf p}_{\rm trig};{\bf p}_j - \sum_{n=1}^N {\bf q}_n
- {\bf \Delta}_r ).
\end{eqnarray}
Using this definition, the momentum distribution of the ridge particle
momentum distribution per trigger particle is then the expectation
value of the sum of the final momentum distribution of the kicked
medium partons:
\begin{eqnarray}
\label{eq3}
\frac{1}{N_{\rm trig}}
\frac{dN_{\rm ridge}^{AA}}{d{\bf p}}
=\left \langle 
 \frac {2}{3} \sum_{n = 1}^N  f_{Rn} \frac{dF_n}{d{\bf p}}({\bf q}_n)  
\right \rangle. \end{eqnarray}
The above equation can be re-written as
\begin{eqnarray}
\label{full3}
\frac{1}{N_{\rm trig}} \frac{dN_{\rm ridge}^{AA}}{d{\bf p}} 
= \frac{2}{3}
\left \{ \langle f_R \rangle \langle N \rangle  \langle 
\frac{dF}{d{\bf p}} \rangle 
+ \left \langle \sum_{n = 1}^N (f_{Rn}- \langle f_R \rangle) 
  [ \frac{dF_n}{d{\bf p}}({\bf q}_n) - \langle
\frac{dF}{d{\bf p}} \rangle ] \right \rangle \right \},
\end{eqnarray}
where $f_R$ is the average attenuation factor,
$f_R=\sum_{n=1}^{N}f_{nR}/N$, and $\langle N \rangle$ is the
expectation value of the total number of kicked medium partons per
trigger particle.  The quantity $\langle N\rangle$ is also the
expectation value of the number of jet-(medium parton) collisions per
trigger particle.  We shall often label $\langle N\rangle$
alternatively as $\langle N_k\rangle$ with the subscript $k$ to
emphasize that this is the averaged number of $k$icked medium partons
per trigger.

As defined by Eq.\ (\ref{expt}), $\langle N \rangle$ and $\langle
N_k \rangle$ are given by
\begin{eqnarray}
\label{avern}
{\langle N \rangle} \equiv {\langle N_k \rangle} = 
 \frac{1}{N_{\rm trig}}
\int {d{\bf p}_j}
\frac{dN_j}{d{\bf p}_j} \sum_{N=0}^{N_{\rm max}}
NP(N) e^{-\zeta_a N} 
D({\bf p}_{\rm trig};{\bf p}_j - \sum_{n=1}^N {\bf q}_n
- {\bf \Delta}_r
).
\end{eqnarray}
To understand the gross features of the phenomenon, we neglect the
second term in the curly bracket of Eq. (\ref{full3}) which arises
from the fluctuation of the quantities from their mean values.  The
formulation can also be simplified by taking the different momentum
kicks ${\bf q}_n$ to be the average ${\bf q}$.  Using these
simplifying assumptions, we then obtain
\begin{eqnarray}
\label{eq11}
\frac{1}{N_{\rm trig}} \frac{dN_{\rm ridge}^{AA}}{d{\bf p}} = \langle
 f_R \rangle \frac{2}{3} \langle   N_{\rm k} \rangle \langle \frac{dF}{d{\bf
 p} } \rangle.
\end{eqnarray}
Thus, the ridge particle distribution is separated into a
geometry-dependent part $\langle f_R\rangle (2/3){\langle N_{k}
\rangle}$ and the average normalized momentum distribution of ridge
particles, $\langle dF/d{\bf p} \rangle $.  For brevity of notation,
the bracket symbol, $\langle \rangle$, for $\langle dF/d{\bf p}
\rangle $ will be made implicit, and the normalized ridge momentum
distribution $dF/d{\bf p}$ will be understood to represent the average
over the jet source distribution, jet collision
locations, and jet energies. 

If one is interested in the total ridge yield by integrating over the
ridge particle momentum, we then get
\begin{eqnarray}
\label{eq12}
\frac{N_{\rm ridge}^{AA}}{N_{\rm trig}}
=\langle f_R  \rangle \frac{2}{3}  \langle N_{k} \rangle . 
\end{eqnarray}

Our strategy is to study first the case of the most-central AuAu
collisions at $\sqrt{s_{NN}}=200$ GeV where the momentum distribution
of the ridge particles and the average number of kicked medium partons
can be inferred from experimental data \cite{Ada05,Put07,Wan07}.  In
Sections XI, XII, and XIII, we shall then examine the average number
of kicked medium partons and the experimental ridge yield as a
function of centrality, collision energies, and nuclear mass numbers
\cite{Put07,Bie07,Nat08}, using the number of kicked medium partons
for the most-central AuAu collision at $\sqrt{s_{NN}}=200$ GeV as a
reference.

\section{Relation between the initial and final momentum distributions}

In the momentum kick model, the normalized final parton momentum
distribution $E dF/d{\bf p}$ at ${\bf p}$ is related to the normalized
initial parton momentum distribution $E_i dF/d{\bf p}_i$ at ${\bf
p}_i$ at a shifted momentum, ${\bf p}_i={\bf p}-{\bf q}$, and we have
\cite{Won07}
\begin{eqnarray}
\label{final}
\frac{dF}{ p_{t}dp_{t}d\eta d\phi} &=&\left [ \frac{dF}{
p_{ti}dp_{ti} dy_i d\phi_i } \frac{E}{E_i} \right ]_{{\bf p}_i
={\bf p}-{\bf q}}
\sqrt{1-\frac{m^2}{(m^2+p_t^2) \cosh^2 y}},
\end{eqnarray}
where the factor $E/E_i$ insures conservation of particle numbers and
the last factor changes the rapidity distribution to the
pseudorapidity distribution \cite{Won94}.  The momentum kick ${\bf q}$
is expected to lie within a narrow cone in the trigger particle
direction for a high-energy jet.  To minimize the number of
parameters, we approximate ${\bf q}$ to lie along the trigger particle
direction.

To relate the final parton momentum distribution to the observed
hadron momentum distribution, we assume hadron-parton duality which is
a reasonable description for the hadronization of energetic partons.
The final parton momentum distribution Eq.\ (\ref{final}), multiplied
by $\langle f_R \rangle (2/3) \langle N_{k}\rangle$, can then be
identified with the observed (charged) hadron associated particle
momentum distribution per trigger, $dN_{\rm ch}/N_{\rm trig} d\eta
d\phi p_tdp_t$, as given by Eq.\ (\ref{eq11}).  By a simple change of
variables, we can further obtain $dN_{\rm ch}/N_{\rm trig} d\Delta
\eta d\Delta \phi p_t dp_t$ in terms of $\Delta \eta=\eta - \eta_{\rm
jet}$ and $\Delta \phi = \phi-\phi_{\rm jet}$, relative to the trigger
particle.  The basic ingredients of the momentum kick model are then
the magnitude of the momentum kick $q$, the normalized initial parton
momentum distribution $dF/d{\bf p}_i$, and the average number of
jet-(medium parton) collisions per jet.  For numerical calculations,
we set $m=m_\pi$.

The initial and final parton momenta can be represented in terms of
Cartesian components in the collider frame,
${\bf  p}=(p_{1},p_{2},p_{3})$, with a longitudinal $p_3$ component, a
transverse $p_1$ component, and another transverse $p_2$ component
perpendicular to both $p_1$ and $p_3$.  The coordinate axes can be so
chosen that the trigger jet lies in the $p_1$-$p_3$ plane.  The
initial parton momentum ${\bf p}_i=(p_{i1},p_{i2},p_{i3})$ is related
to the final momentum ${\bf p}_f=(p_{f1},p_{f2},p_{f3})$ and the
trigger jet rapidity $\eta_{\rm jet}$ by
\begin{subequations}
\begin{eqnarray}
p_{i1}&=&p_{f1}-\frac{q}{\cosh \eta_{\rm jet}},\\
p_{i2}&=&p_{f2},\\
p_{i3}&=&p_{f3}-\frac{q\sinh \eta_{\rm jet}}{\cosh \eta_{\rm jet}}.
\end{eqnarray}
\end{subequations}
For a given trigger particle pseudorapidity, these relations allow one
to obtain ${\bf p}_i$ from ${\bf p}_f={\bf p}$ for the evaluation of
the ridge yield per trigger particle.

\section{Parametrization of the Initial Parton Momentum Distribution}

As the jet-(medium parton) collisions take place at different spatial
and temporal locations during the passage of the near-side jet through
the medium, the initial momentum distribution $E_i dF/d{\bf p}_i$ in Eq.\
(\ref{final}) refers actually to an average over spatial and temporal
regions during the early stage of the nucleus-nucleus collision.  The
`initial' momentum distribution can also be called the `early' parton
momentum distribution.  This initial parton momentum distribution
$E_i dF/d{\bf p}_i$ of the medium partons at the time of jet-(medium
parton) collisions is not yet a quantity that can be obtained from
first principles of QCD, although some of its qualitative features can
be inferred from basic principles as will be discussed in Sections
VIII and IX.  Furthermore, jets occur at an early stage of the
nucleus-nucleus collisions, whereas the momentum distribution of the
bulk medium pertains to the state of the medium at the end-point of
the nucleus-nucleus collision.  Therefore the early parton momentum
distribution near the beginning stage of the nucleus-nucleus collision
needs not be the same as that of the bulk matter at the end-point of
the nucleus-nucleus collision.

Under the circumstances, the parton momentum distribution at the early
stage of the nucleus-nucleus collision can only be obtained
phenomenologically from the ridge particle data by representing it as
a simple function whose distinct characteristics can be determined by
comparison with experimental ridge data.  

The initial momentum distribution was parametrized previously as
$e^{-y_i^2/2\sigma_y^2}
\exp\{-\sqrt{m^2+p_{ti}^2}/T\}/\sqrt{m^2+p_{ti}^2}$, with $m$ taken to
be $m_\pi$ \cite{Won07}.  Although this is adequate for mid-rapidity
and high-$p_t$ ridge particles \cite{Won07}, it leads to too large a
ridge yield both at $p_{t}\sim 1$ GeV (dotted curve in Fig. 2) and at
forward rapidities.  Our understanding of the behavior of the early parton
transverse momentum distribution has not reached such a stage that we
can predict its low $p_t$ behavior definitively.  If the partons arise
from a deconfined medium with a finite transverse boundary, then the
transverse parton momentum distribution at small $p_t$ will be
flattened from an exponential distribution, as shown in Figs. 1 and 2
of \cite{Mos95}.  Transverse distributions of this type can be
described by replacing the denominator $\sqrt{m^2+p_{ti}^2}$ with
$\sqrt{m_d^2+p_{ti}^2}$ where $m_d$ can be adjusted to lead to the
correct ridge yield at $p_{t}\sim 1$ GeV.  The extracted transverse
momentum distribution may provide useful information to study the
transverse radius of the deconfined parton medium \cite{Won93,Mos95}.

\begin{figure} [h]
\includegraphics[angle=0,scale=0.40]{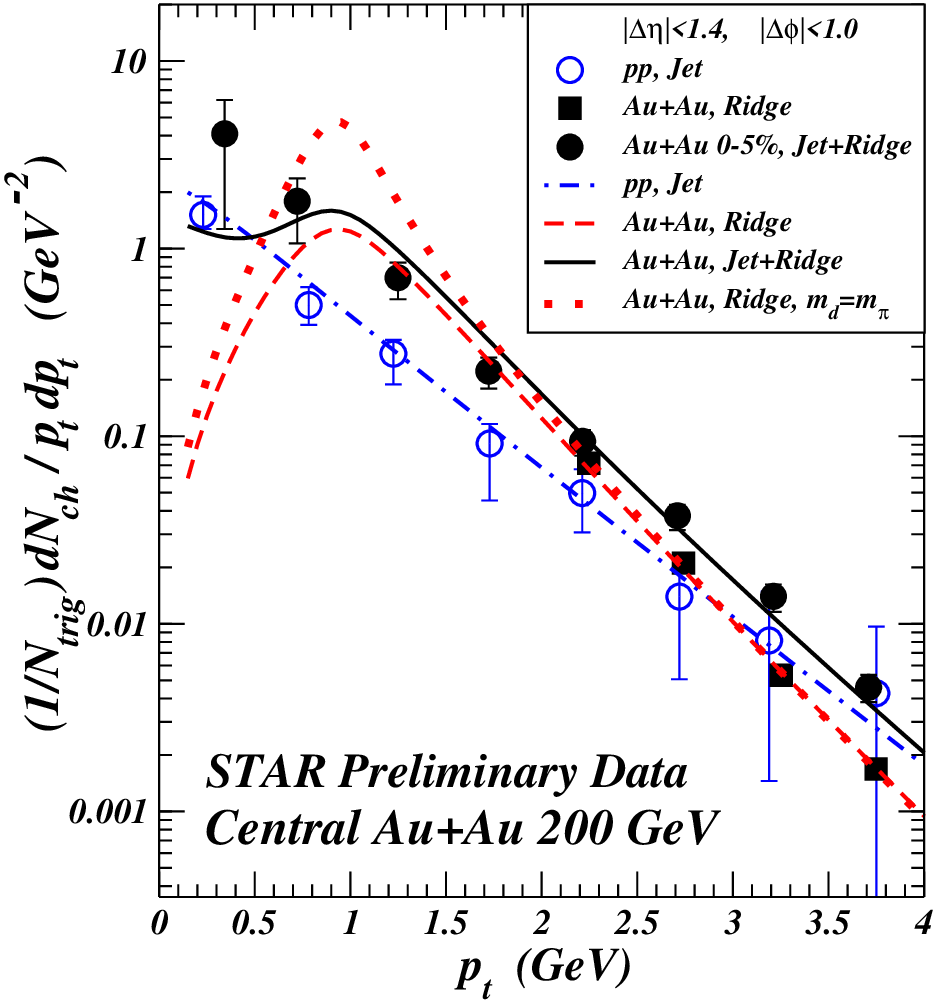}
\vspace*{0.0cm}
\caption{ (Color online) The symbols represent STAR experimental data
\cite{Ada05,Put07} and the curves theoretical results of $dN_{\rm
ch}/N_{\rm trig}p_t dp_t$, for $pp$ and central AuAu collisions.}
\end{figure}

The difficulty with the forward rapidity region can be resolved by
noting that the Gaussian rapidity distribution of \cite{Won07} does
not take into account the kinematic boundary restrictions on phase
space.  The large values of $\sigma_y$ extracted from the mid-rapidity
data in \cite{Won07} imply that the rapidity distribution is quite
flat in the mid-rapidity region.  We can use a rapidity distribution
that retains the flatness at mid-rapidity but also respects the
kinematic boundaries at large rapidities and large $p_t$. Accordingly,
we parametrize the normalized initial parton momentum distribution as
\begin{eqnarray}
\label{dis2}
\frac{dF}{ p_{ti}dp_{ti}dy_i d\phi_i}&=&
A_{\rm ridge} (1-x)^a 
\frac{ e^ { -\sqrt{m^2+p_{ti}^2}/T }} {\sqrt{m_d^2+p_{ti}^2}},
\end{eqnarray}
where $A_{\rm ridge}$ is a normalization constant defined (and
determined numerically) by
\begin{eqnarray}
\int dy_i d\phi_i p_{ti}dp_{ti}
{A}_{\rm ridge} (1-x)^a 
\frac{ \exp \{ -\sqrt{m^2+p_{ti}^2}/T \}} {\sqrt{m_d^2+p_{ti}^2}}
= 1.
\end{eqnarray} 
In Eq.\ (\ref{dis2}), $x$ is the light-cone variable \cite{Won94}
\begin{eqnarray}
\label{xxx}
x=\frac{\sqrt{m^2+p_{ti}^2}}{m_b}e^{|y_i|-y_b},
\end{eqnarray}
$a$ is the fall-off parameter that specifies the rate of decrease of
the distribution as $x$ approaches unity, $y_b$ is the beam parton
rapidity, and $m_b$ is the mass of the beam parton whose collision and
separation lead to the inside-outside cascade picture of particle
production \cite{Cas74,Bjo83,Won91,Won94,And83}.  As $x \le 1$, there
is a kinematic boundary that is a function of $y_i$ and $p_{ti}$,
\begin{eqnarray}
\label{pty}
\sqrt{m^2+p_{ti}^2}=m_b e^{y_b-|y_i|}.
\end{eqnarray}
We expect $y_b$ to have a distribution centered around the nucleon
rapidity, $y_N=\cosh^{-1}(\sqrt{s_{_{NN}}}/2m_N)$.  For lack of a
definitive determination, we shall set $y_b$ equal to $y_N$ and $m_b$
equal to $m_\pi$, pending their future experimental determination by
examining the ridge boundaries.  This form of the initial parton
distribution leads to a restricted phase space that is smaller than
that for a Gaussian rapidity distribution.  As a consequence, it can
lead to a smaller associated particle yield that agrees with
experimental forward rapidity data as shown in Section VI.

\section{Particle Momentum Distribution of the Jet Component }

As a jet passes through the parton medium, the medium partons kicked
by the jet will materialize to become particles in the associated
``ridge component'', while the jet will fragment and radiate into the
trigger and associated ``jet component'' particles.  If the
contribution from the jet component is known, we can separate out the
ridge component using experimental data of total associated particles.
The distribution of the ``jet component'' of (charged) associated
fragmentation products is given by
\begin{eqnarray}
\label{jet1}
\frac{dN_{\rm jet}^{AA}}{d{\bf p}}
=\int d{\bf p}_j \frac{dN_j}{d{\bf p}_j} \sum_{N=0}^{N_{\rm max}} 
P(N)  e^{-\zeta_a N} 
D_2({\bf
p}_{\rm trig},{\bf p};{\bf p}_j - \sum_{n=1}^{N} {\bf q}_n
-{\bf \Delta}_r
),
\end{eqnarray}
where $D_2({\bf p}_a, {\bf p}_b ; {\bf p}_c)$ is the double
fragmentation function for fragmenting into hadrons of momentum ${\bf
  p}_a$ and ${\bf p}_b$ from a jet parton of momentum ${\bf
  p}_c$.  Fragmentation measurements \cite{Put07} suggest an
approximate scaling relation
\begin{eqnarray}
D_2({\bf p}_{\rm trig}, {\bf p}; {\bf p}_c)
\approx D({\bf p}_{\rm trig}; {\bf p}_{c}) 
D_z({\bf p};{\bf p}_{\rm trig}),
\end{eqnarray}
where $D_z({\bf p} ; {\bf p}_{\rm trig} )$ is approximately the same
(within a factor of about 0.6 to 1.2) for dAu and AuAu collisions in
$2.5 <p_{t,\rm trig}< 6$ GeV (Fig. 5b and 5c of \cite{Put07}).
By applying this approximate scaling relation to Eq. (\ref{jet1}) and
using Eq.\ (\ref{trig}), the (charged) jet component in an $AA$ central
collision per trigger is
\begin{eqnarray}
\label{jet}
\frac{1}{N_{\rm trig}} \frac {dN_{\rm jet}^{AA}}{d{\bf p}} \approx
D_z({\bf p};{\rm p}_{\rm trig}) \approx \frac{dN_{\rm jet }^{pp}}{d{\bf
p}}.
\end{eqnarray}
Because of the approximate nature of the above relation (up to a
factor of about 0.6 to 1.2), we need to make a quantitative check. In
the region where the jet component has a prominent appearance, as in
Fig.\ 3(d) for $2< p_t < 4 $ GeV, the total $dN_{\rm ch}/N_{\rm
trig}d\Delta \eta$ distribution at $\Delta\eta\sim 0$ has indeed a
shape similar to, but a peak magnitude about equal to, the $pp$
near-side jet distribution.  The total yield is the sum of the
jet component and the ridge yield, and the ridge yield at $\Delta\eta\sim
0$ is non-zero and nearly flat (Fig. 3(d)). The near-side jet
component in AuAu central collisions per trigger is thus an attenuated
$pp$ near-side jet distribution, as expected for production in an
interacting medium. If one assumes that fragmentation products lying
deeper than an absorption length from the surface are all absorbed,
then the average jet fragment attenuation factor is
$f_J=\int_0^\lambda e^{-x/\lambda}dx/\lambda = 0.632$, which leads
semi-empirically to a reasonable description of the experimental data
as indicated below in Figs. 2 and 3.

\begin{figure} [h]
\includegraphics[angle=0,scale=0.50]{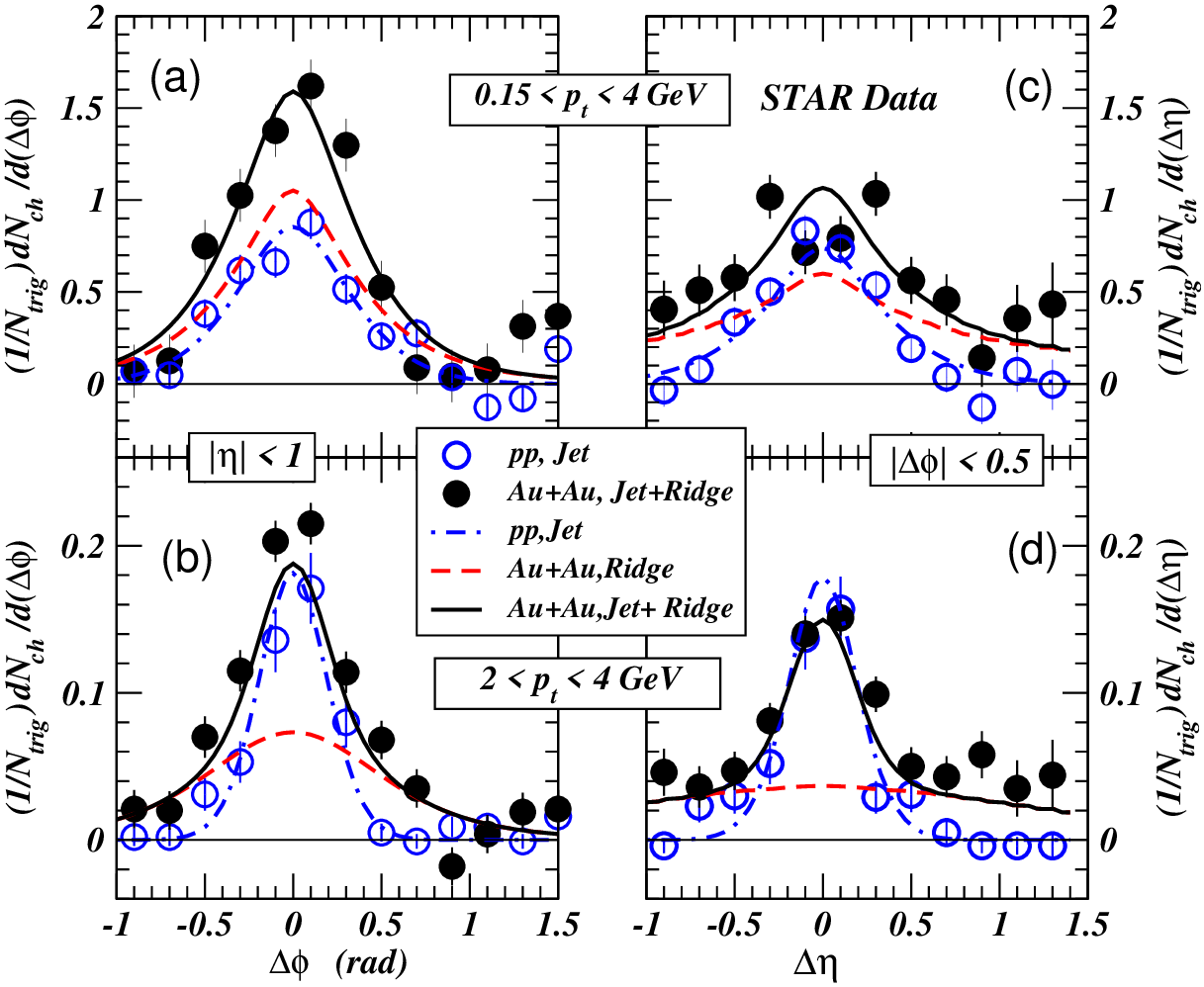}
\vspace*{0.0cm} 
\caption{ (Color online) The symbols represent experimental data
 \cite{Ada05} and the curves theoretical results, for $pp$ and central
 AuAu collisions.  (a) and (b) give the $dN_{\rm ch}/N_{\rm
 trig}d\Delta \phi$ distributions.  (c) and (d) give the $dN_{\rm
 ch}/N_{\rm trig}d\Delta \eta$ distributions.  }
\end{figure}

The sum of the distributions (\ref{eq11}) and (\ref{jet}), relative
to the trigger particle $\eta_{\rm jet}$ and $\phi_{\rm jet}$, is
therefore given more precisely as
\begin{eqnarray}
\label{obs}
\left [ 
\frac{1}{N_{\rm trig}}
\frac{dN_{\rm ch}} 
{p_{t} dp_{t} d\Delta \eta  d\Delta \phi } \right ]_{\rm total}^{\rm AA} 
= \left [ \langle f_R \rangle  \frac {2}{3} 
\langle N_k \rangle \frac { dF } {p_t dp_t\,
d\Delta \eta\, d\Delta \phi} \right ]_{\rm ridge}^{\rm AA} 
+
\left [ f_J  \frac { dN_{\rm jet}^{pp}} {p_t dp_t\, d\Delta \eta\, d\Delta
\phi} \right ]_{\rm jet}^{\rm AA} \!\!\!\!.
\end{eqnarray}

The experimental momentum distribution of (charged) near-side
particles associated with the trigger in a $pp$ collision, measured
relative to the trigger jet, can be parametrized as
\begin{eqnarray}
\label{jetfun}
\frac { dN_{\rm jet}^{pp}} {p_t dp_t\, d\Delta \eta\, d\Delta \phi}
\!\!= N_{\rm jet}
\frac{\exp\{(m-\sqrt{m^2+p_t^2})/T_{\rm jet}\}} {T_{\rm jet}(m+T_{\rm jet})}
\frac{1}{2\pi\sigma_\phi^2}
e^{- {[(\Delta \phi)^2+(\Delta \eta)^2]}/{2\sigma_\phi^2} },
\end{eqnarray}
where $N_{\rm jet}$ is the number of (charged) near-side jet particles
in a $pp$ collision, and $T_{\rm jet}$ is the jet inverse slope
(``temperature'') parameter. The above functional form of the jet
fragmentation product cone in terms of $\Delta \phi$ and $\Delta \eta$
was chosen because $p_2=p_t\sin \Delta \phi$ and $p_3=p_t \sinh \Delta
\eta$, and the square of the momentum perpendicular to the jet
direction has a magnitude
\begin{eqnarray}
p_2^2+p_3^2= p_t^2 \sin^2 \Delta \phi+ p_t^2 \sinh^2 \Delta \eta \sim
p_t^2 [ (\Delta \phi)^2 + (\Delta \eta)^2] ~~{\rm for~small~}\Delta
\phi~{\rm and~} \Delta \eta.
\end{eqnarray}
The above equation indicates the symmetry between $\Delta \phi$ and
$\Delta \eta$ for a narrow jet cone. In this functional form of Eq.\
(\ref{jetfun}) for the jet cone, the width in $\Delta \eta$ is equal to
the width in $\Delta \phi$.

In our search for parameter values we find that the width parameter
$\sigma_\phi$ depends slightly on $p_t$ which we parametrize as
\begin{eqnarray}
\label{ma}
\sigma_\phi=\sigma_{\phi 0} \frac{m_a}{\sqrt{m_a^2+p_t^2}}.
\end{eqnarray}

Experimental data for near-side jet particles in $pp$ and central AuAu
collisions obtained by the STAR Collaboration, within the detector
acceptance of $|\eta_{\rm associated}|<1$ and $|\eta_{\rm jet}|<0.7$,
are given in Figs.\ 2 and 3 \cite{Ada05}.  Figure 2 gives the $dN_{\rm
ch}/N_{\rm trig} p_t dp_t$ data, obtained by integrating $dN_{\rm
ch}/N_{\rm trig} p_t dp_t d\Delta \phi d\Delta \eta$ over the domain
of $|\Delta \eta |< 1.4$ and $|\Delta \phi | < 1.0$.  Figures 3(a) and
3(b) give $dN_{\rm ch}/N_{\rm trig} d\Delta \phi$ data, and Figures
3(c) and 3(d) give $dN_{\rm ch}/N_{\rm trig}d\Delta \eta$ data.  They
are obtained by integrating $dN_{\rm ch}/N_{\rm trig}p_t dp_t d\Delta
\phi d\Delta \eta$ over the domains indicated in the figures.
Specifically, Fig.\ 3(a) covers the domain of $|\eta|<1$ and $0.15<
p_t < 4$ GeV, 3(b) the domain of $|\eta|<1$ and $2< p_t < 4$, 3(c) the
domain of $|\Delta \phi|<0.5$ and $0.15< p_t < 4$ GeV, and finally
3(d) the domain of $|\Delta \phi|<0.5$ and $2< p_t < 4$ GeV.  The
domains of integration in a $pp$ collision and a nucleus-nucleus
collision are the same, and the distribution in $\Delta \eta$ has been
corrected for detector acceptance.

The set of experimental $pp$ near-side jet data of $dN_{\rm
  ch}^{pp}/p_t dp_t$, $dN_{\rm ch}^{pp}/ d\Delta \phi$, and $dN_{\rm
  ch}^{pp} /d\Delta \eta$, represented by open circle points in
Figs. 2 and 3, can be described by Eqs.\ (\ref{jetfun}) and
(\ref{ma}), with the following parameters
\begin{eqnarray}
\label{jetpar}
T_{\rm jet}=0.55{\rm ~GeV}, ~\sigma_{\phi 0}=0.50,
~m_a=1.1 {\rm ~GeV,~and~}
N_{\rm jet}=0.75. 
\end{eqnarray}
Theoretical $pp$ jet results obtained with this set of parameters
within the specified experimental domain are shown as the dash-dot
curves in Figs.\ 2 and 3.  They yield a reasonable description of the
experimental momentum distributions of jet particles associated with
the near-side jet in a $pp$ collision.

\section{Comparison of Theoretical Near-Side Associated Particle 
Yields with Experiment}

Theoretical evaluation of both the jet component and the ridge
component for central AuAu collisions allows one to determine the
total yield of associated particles as determined by Eq.\ (\ref{obs}).
A self-consistent search for the initial parton momentum distribution
in Eqs.\ (\ref{dis2}) and (\ref{final}) can be made by comparing the
momentum kick model results with experimental data for $dN_{\rm
ch}/N_{\rm trig} p_t dp_t$, $dN_{\rm ch}/N_{\rm trig} d\Delta \phi$,
and $dN_{\rm ch}/N_{\rm trig} d\Delta \eta$ for mid-rapidities in
Figs.\ 2 and 3, and $dN_{\rm ch}/N_{\rm trig} d\Delta \phi$ for
forward rapidities in Fig.\ 4.  We find that the totality of the STAR
associated particle data, from $p_t=0.15$ GeV to 4 GeV and $\eta$ from
zero up to 3.9 in central AuAu collisions at $\sqrt{s_{NN}}=200$ GeV
\cite{Ada06,Put07,Wan07}, can be described by
Eqs.\ (\ref{obs}) and (\ref{final}) with parameters
\begin{eqnarray}
\label{bestset}
q=1.0 {\rm ~GeV}, {\rm ~~and~~} \langle f_R \rangle \langle N_k
\rangle =3.8,
\end{eqnarray}
in conjunction with the initial parton momentum distribution Eq.\
(\ref{dis2}) with parameters
\begin{eqnarray}
\label{bestset1}
~T=0.50 {\rm ~GeV}, ~m_d=1 {\rm ~GeV},
~{\rm ~~and~~} a=0.5.
\end{eqnarray}

We now discuss the comparison of the experimental data with
theoretical results involved in the above analysis.  In Fig. 2 the
STAR experimental $dN_{\rm ch}/N_{\rm trig}p_t dp_t$ data
\cite{Ada05,Put07} are represented by solid circular points for
central AuAu collisions and by open circle points for $pp$ collisions.
The theoretical results for $pp$ collisions obtained with the
parametrization of Eq.\ (\ref{jetfun}) with parameters in
Eq.\ (\ref{jetpar}) are shown as the dash-dot curves, which agree with
$pp$ near-side data in all experimental kinematic regions.
Experimental ridge $dN_{\rm ch}/N_{\rm trig}p_tdp_t$ data in central
AuAu collisions \cite{Put07} are also shown as solid squares and they
are calibrated by using the data of Fig.\ 2 of \cite{Put07}.  The
solid curve is the theoretical result for $dN_{\rm ch}/N_{\rm trig}p_t
dp_t$ for central AuAu collisions, as the sum of the jet part and the
ridge part, with the ridge part of the contribution shown as the
dashed curve.  They have been calculated with $m_d=1$ GeV.  If we set
$m_d$ equal to $m_\pi$, then we will get the ridge yield represented
by the dotted curve, which over-predicts the ridge yield at $p_t\sim
1$ GeV.
 
Fig.\ 2 shows good agreement between theoretical $dN_{\rm ch}/N_{\rm
trig}p_t dp_t$ with experimental data for central AuAu collisions.
The theoretical transverse momentum distribution of the jet and the
ridge components have very different shapes in the low $p_t$ region.
The jet component $dN_{\rm ch}/N_{\rm trig}p_t dp_t$ decreases
exponentially as a function of increasing $p_t$.  On the other hand,
the magnitude of the final transverse parton momentum $p_{tf}$ is
greater than the initial transverse momentum $p_{ti}$ approximately by
the amount $q$.  The initial momentum distribution $dF/p_{ti} dp_{ti}$
has a peak at $p_{ti}=0$.  As a consequence, the theoretical ridge
yield of final partons, $dN_f/N_{\rm trig}p_{tf} dp_{tf}$, has a peak
around $p_{tf}\sim q \sim 1$ GeV and it decreases significantly for
small values of $p_t$, in contrast to the exponential behavior of the
jet component.

It is interesting to note that the theoretical ratio of the jet yield
to the ridge yield is greater than 1 for $p_t \lesssim 0.6$ GeV, but
is less than 1 in the interval $0.6 \lesssim p_t \lesssim 3.7$ GeV.
The ratio reverts to become greater than 1 at $3.7 {\rm ~GeV}\lesssim
p_t $.  The change of the dominance of the ridge component as $p_t$
changes may lead to experimentally observable variations of the shape
of the $dN_{\rm ch}/N_{\rm trig}d\Delta \eta$ as a function of $p_t$.

\begin{figure} [h]
\includegraphics[angle=0,scale=0.40]{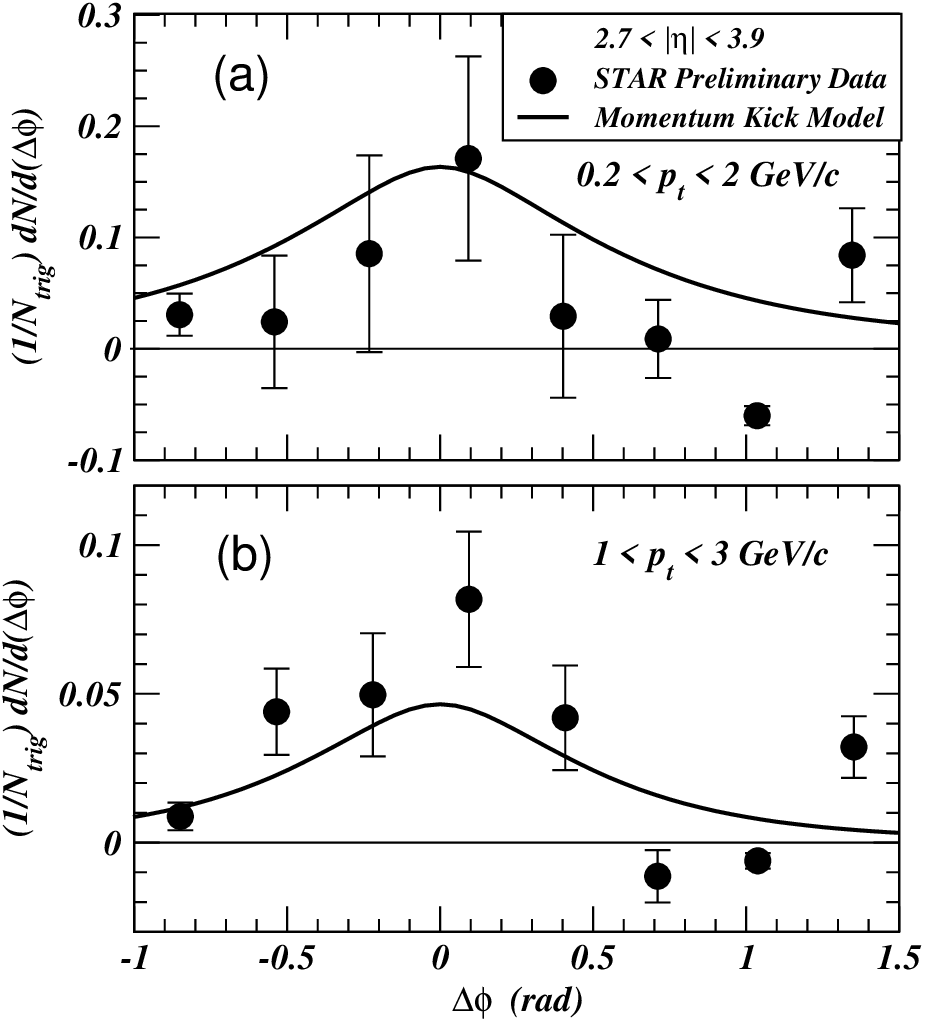}
\vspace*{0.0cm}
\caption{ (Color online) Azimuthal angular distribution data at
forward pseudorapidities for central AuAu collisions from the STAR
Collaboration \cite{Wan07}, compared with theoretical results shown as
solid curves from the momentum kick model.  (a) is for $0.20 < p_t <
2$ GeV, and (b) is for $1 < p_t < 3$ GeV.  }
\end{figure}

In Fig.\ 3, the experimental total associated particle yields
\cite{Ada05,Wan07} are represented by solid circular points for
central AuAu collisions and by open circles for $pp$ collisions.  The
theoretical results for $pp$ collisions obtained with the
parametrization of Eq.\ (\ref{jetfun}) are shown as the dash-dot
curves which agree with experimental $pp$ near-side data.  In
these figures, the theoretical total yield and the ridge yield for
central AuAu collisions are represented by solid and dashed curves,
respectively.  Comparison of the theoretical total yield and the
experimental total associated particle yield for central AuAu
collisions indicates general agreement over all azimuthal angles
[Figs. 3(a) and 3(b)] and over all pseudorapidities [Figs. 3(c) and
3(d)], for both $0.15 < p_t < 4 $ GeV [Figs. 3(a) and 3(c)] and $2 <
p_t < 4 $ GeV [Figs. 3(b) and 3(d)].

One notes from Fig.\ 3(a) that for the region of $0.15 < p_t < 4$ GeV,
which receives the dominant contributions from the low $p_t$ region,
the widths of the azimuthal angular distributions of the ridge and the
jet components are nearly the same, with the magnitude of the ridge
yield slightly higher than the $pp$ yield.  On the other hand, in the
region of $2<p_t<4$ GeV in Fig.\ 3(b), which receives the dominant
contributions from the region near $p_t\sim 2$ GeV, the azimuthal
angular distributions of the jet component is narrower than the ridge
component azimuthal angular distribution.

We observe from Fig.\ 3(c) that the theoretical AuAu jet and ridge
components have different shapes in $dN_{\rm ch}/N_{\rm trig}d\Delta
\eta$.  The jet component maintains a sharp peak in $dN_{\rm
ch}/N_{\rm trig}d\Delta \eta$.  In the low $p_t$ region, the
pseudorapidity distribution of the theoretical ridge component is
significantly broader than the jet component and its magnitude remains
to have a non-zero value at large $|\Delta \eta|$.  In the high $p_t$
region in Fig. 3(d), the ridge pseudorapidity distribution is
essentially flat and non-zero.  The broad peak structure for the low
$p_t$ region comes from the factor $E/E_i$ in Eq.\ (\ref{final}),
arising from the difference of the momenta of the parton before and
after the collision.  This factor is close to 1 for the high $p_t$
region, and the flatness of the distribution is a reflection of the
initial rapidity distribution.

We turn now to forward rapidities where preliminary experimental data
for central AuAu collisions have been obtained for $2.7 <|\eta|< 3.9$
\cite{Wan07}. We note that $dN_{\rm ch}/N_{\rm trig}d\Delta \phi
d\Delta \eta $ at $\Delta \phi \sim 0$ for $|\eta| < 1 $ in Fig.\ 3(a)
is an order of magnitude greater than the corresponding $dN_{\rm
ch}/N_{\rm trig}d\Delta \phi d\Delta \eta $ for 2.7$<$$|\eta|$$<$3.9
in Fig.\ 4(a). This implies a substantial fall-off of the ridge yield
$dN_{\rm ch}/N_{\rm trig}\Delta \phi d\Delta \eta$ at $\Delta
\phi$$\sim$0 in going from mid-rapidities to large rapidities.  Even
though the mid-rapidity data place a constraint on the flatness of the
mid-rapidity distribution, they do not otherwise constrain the rate of
fall-off of the distribution in the forward rapidity region.
Measurements at forward rapidities in Fig. 4 contain events with large
$\eta$ and $p_t$ that are either already outside the kinematic limits
or close to the kinematic limits.  Therefore, even with large
errors, the forward rapidity data in Fig. 4 are sensitive to the
constraint of the kinematic limits and the rate of fall-off of the
initial parton momentum distribution as specified by the fall-off
parameter $a$.  The data of Fig.\ 4 lead to $a=0.5$.

\begin{figure} [h]
\includegraphics[angle=0,scale=0.40]{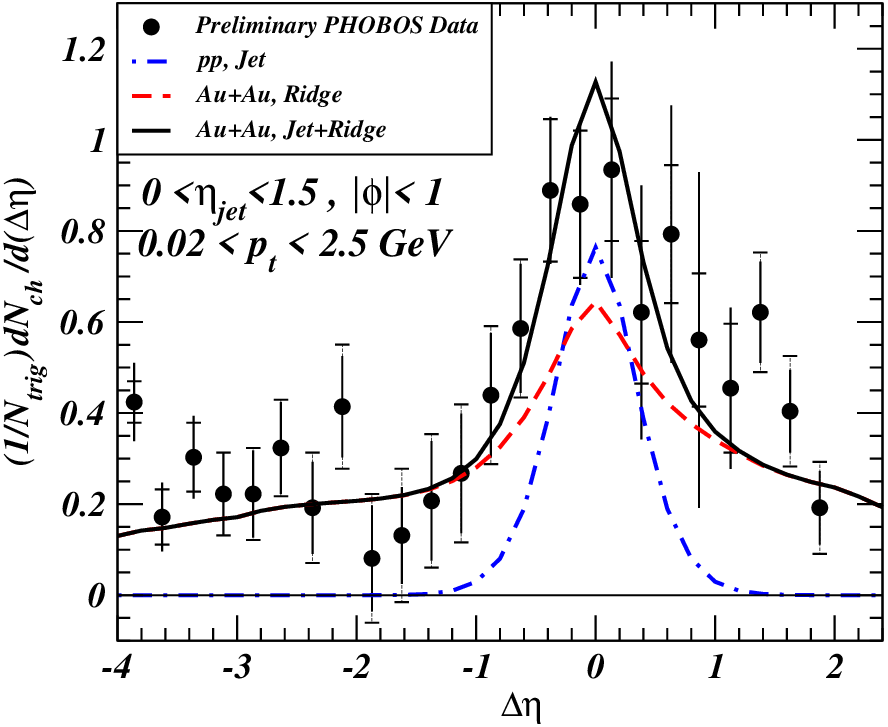}
\vspace*{0.0cm}
\caption{ (Color online) The momentum distribution of the associated
particles as a function of the pseudorapidity relative to the jet
pseudorapidity $\Delta \eta$. The solid circular points are the data
from the PHOBOS Collaboration \cite{Wen08} and the curves give
theoretical predictions from the momentum kick model.  The solid,
dashed, and dash-dot curves give the total yield, the ridge yield, and
the $pp$ jet yield, respectively.  }
\end{figure}

Using the parameters we have extracted from the STAR ridge data as
given by (\ref{bestset}) and (\ref{bestset1}), we can predict the
pseudorapidity distribution for the PHOBOS experimental acceptance
defined by $\Delta \phi \le 1$, $0<\eta_{\rm trig}<1.5$, and
$0.02<p_t<2.5$ GeV.  The theoretical total associated particle yield,
which is the sum of the ridge yield and the attenuated $pp$ jet yield,
is shown as the solid curve in Fig. 5.  The theoretical $pp$ jet yield
and the ridge component of the associated particles are shown as the
dash-dot and the dashed curves respectively.  The result has been
corrected for $\Delta \eta$ acceptance.  The present prediction of the
momentum kick model for the near-side jet associated particle yields
was found to agree well with experimental measurements obtained by the
PHOBOS Collaboration \cite{Wen08} up to large $|\Delta \eta|$ for the
region of small $p_t$.

\section{Extracted Initial  Parton Momentum Distribution}
 
It is illuminating to examine the initial parton momentum distribution
extracted from the totality of experimental data in Figs.\ 2, 3, and
4.  We find that the normalized initial parton momentum distribution
at the jet-(medium parton) collisions can be represented by $dF/dy
d\phi p_t dp_t =A_{\rm ridge} (1-x)^a
\exp\{-\sqrt{m^2+p_t^2})/T\}/\sqrt{m_d^2+p_t^2}$, where
$x=\sqrt{m^2+p_t^2}e^{|y|-y_b}/m_b$, $a$=0.5, $T=0.5$ GeV, and $m_d=1$
GeV.  Here, $(y,\phi,p_t)$ represent the initial parton momentum
coordinates. We show explicitly the extracted, normalized initial
parton distribution $dF/p_t dp_t dy$ at the moment of jet-(medium
parton) collisions in Fig.\ 6.  It is given as a function of $p_t$ for
various $y$ in Fig. 6(a), and conversely as a function of $y$ for
various $p_t$ in Fig. 6(b).  It has a thermal-like transverse momentum
distribution and is nearly flat in rapidity at $y\sim 0$, with sharp
kinematic boundaries at large $|y|$.

\begin{figure} [h]
\includegraphics[angle=0,scale=0.40]{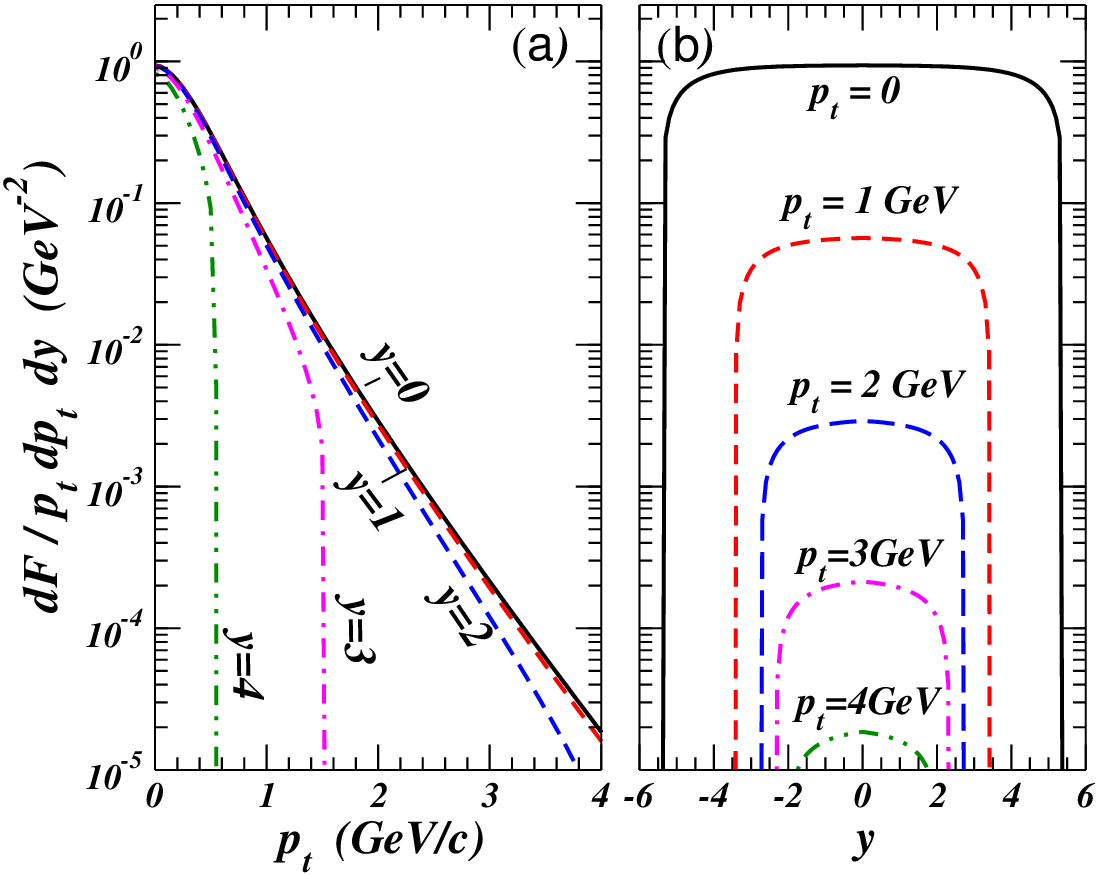}
\vspace*{0.0cm}
\caption{ (Color online) Normalized initial parton momentum
distribution $dF/dy p_t dp_t$ extracted from the STAR Collaboration
data \cite{Ada05,Put07,Wan07}. (a) $dF/dy p_t dp_t$ as a function of
$p_t$ for different $y$, and (b) $dF/dy p_t dp_t$ as a function of $y$
for different $p_t$.  }
\end{figure}

In Fig. 6(a), the momentum distribution for $y=0$ and high $p_t$ has a
slope parameter $T$ that is intermediate between that of the jet and
inclusive bulk particles.  This indicates that partons at the moment
of jet-(medium parton) collision is at an intermediate stage of
dynamical equilibration.

The parton momentum distribution cannot be separated as the product of
two independent distributions.  The momentum distribution as a
function of $p_t$ depends on the rapidity variable $y$ which affects
the boundaries of the distribution.  The distribution as a function of
$p_t$ does not change much for $y$ up to $y=2$.  For $y=3$, the
maximum value of $p_t$ is 1.54 GeV and the distribution changes
significantly as the kinematic boundary is approached.  For $y=4$, the
boundary of $p_t$ is located at 0.55 GeV.  

In Fig. 6(b), the momentum distribution as a function of $y$ for a
fixed $p_t$ is essentially flat near central rapidities and it extends
to a maximum value of $|y|_{\rm max}$ that depends on $p_t$, as given
by Eq.\ (\ref{pty}).  The flat distribution changes rather rapidly as
it approaches the kinematic limits.  The kinematic boundary becomes
more restrictive to cover a smaller allowed region of $y$ as $p_t$
increases.  For example, for $p_t=$ 0, 1, 2, 3, and 4 GeV, the maximum
values of $|y|$ are 5.36, 3.4, 2.7, 2.33, and 2.05, respectively.  The
extracted early rapidity distribution exhibits the feature of a
plateau structure in rapidity. The width of the plateau decreases as
$p_t$ increases.

The locations on the kinematic boundaries in Figs. 6(a) and 6(b)
depend on the value of $y_b$ and $m_b$ which have been taken to be
$y_N$ and $m_\pi$ respectively in the present analysis.  Better
determination of these quantities using the measured locations of the
kinematic boundaries may require more refined measurements of the
ridge momentum distribution in many locations in pseudorapidity space.

\section{Early Particle production at high energies}

The momentum distribution extracted from the near-side ridge data
indicates that the early parton rapidity distribution has a plateau
structure which extends well into the high rapidity region.  The width
of the plateau depends on $p_t$. The greater the value of $p_t$, the
narrower is the width of the plateau.  While the evolution scenario of
the early rapidity distribution has been outlined in Section VIII of
\cite{Won07}, we would like to elaborate in more detail the origin of
the rapidity plateau.

It should be kept in mind that the plateau rapidity structure has been
known in QCD particle production processes both experimentally and
theoretically.  In $e^+$-$e^-$ annihilation experiments, the produced
particles exhibit a rapidity plateau structure
\cite{Aih88,Hof88,Pet88,Abe99,Abr99}.  Many earlier theoretical
investigations of QCD particle production processes give a rapidity
plateau distribution when a quark pulls away from an anti-quark at
high energies \cite{Cas74,Bjo83,Won91,Won94,And83}.  We shall review
here the theoretical basis for the occurrence of the rapidity plateau
in an elementary particle production process.

As an exact solution of particle production at high energies starting
from the first principle of QCD is not available, many
phenomenological models have been presented to describe particle
production in nucleus-nucleus collisions \cite{Mod08}.  Common to many
of these models (such as the Lund Model, the Dual Parton Model, the
Mulitple Collision Model, the ART model, the Lexus Model, the Venus
Model, and the Glasma Model, ...) is the elementary particle
production process of a color charge pulling away from an anti-color
charge at high energies at the early stage of a nucleus-nucleus
collision, and the nucleus-nucleus collision consists of many of these
elementary production processes.

We can single out one of the elementary production processes for
examination and study the particle production process in a model that
has many essential features as those in QCD
\cite{Cas74,Bjo83,Won91,Won94,And83}.  The model of QED2
\cite{Sch62,Low71,Cas74,Col75,Col76} are quantum mechanical systems in
which a neutral boson exists as a non-perturbative bound state, much
as mesons are bound states in QCD.  When a positive and negative
charge pair are separated in such a system, the vacuum is so polarized
that the positive and the negative charges are completely screened, in
a manner similar to the confinement of quarks, in which a quark cannot
be isolated. It was demonstrated by Casher, Kogut and Susskind
\cite{Cas74} in QED2 that the rapidity distribution of produced boson
particles in a system of two oppositely charged fermions separating at
high relative velocities exhibits a plateau rapidity structure.  Such
a rapidity plateau structure of produced particles is indeed observed in
high-energy $e^+-e^-$ annihilation experiments as mentioned above
\cite{Aih88,Hof88,Pet88,Abe99,Abr99}. The quark fragmentation function
obtained from QED2 \cite{Fuj89} agrees with that of Field and Feynman
\cite{Fie78} in their phenomenological treatment of QCD strings. These
desirable properties of confinement, charge screening, the existence
of neutral bound states, and the proper high-energy behavior make it
useful for Casher, Kogut, Susskind \cite{Cas74}, Bjorken \cite{Bjo83}
and many others \cite{Won91,Won94,And83} to infer the rapidity plateau
structure of produced particles when a color charge recede away from
an anti-color charge at high energies.

Previously, a scaling argument was presented to reduce QCD at high
energies to an effective two-dimensional field theory by scaling the
longitudinal and temporal coordinates by $\lambda$ and expand the
action in power of $1/\lambda$ \cite{Ver93}.  We shall try an
alternative approach by using the physical argument of transverse
confinement to establish the connection between QCD and QED2, in order
to study particle production in a quantum mechanical framework.

We consider the elementary particle production process in a flux tube
in a nucleus-nucleus collision as a color charge and an anti-color
charge separate from each other at high energies.  Produced particles
are quanta of the interacting fields.  Depending on the environment
temperature, they can be considered as partons in the environment of a
strongly-coupled quark-gluon plasma and as hadrons in a cold QCD
environment at zero temperature.  The QCD fields inside the tube
consists of the gauge fields and the fermion degrees of freedom.  At
high energies, the gauge fields can be greatly simplified by noting
that the transverse gauge fields $A_x$ and $A_y$ are expected to be
proportional to the fermion source transverse velocities, which are
smaller as compared to the longitudinal velocity in the $z$-direction.
It is reasonable to ignore the transverse gauge fields $A_x$, and
$A_y$ so that $A_\mu=(A_0,0,0,A_z)$ containing only $A_0$ and $A_z$
degrees of freedom confined in the tube.

The fermion sector can also be approximated.  We shall assume that as
a result of the non-perturbative non-Abelian gauge interaction,
transverse confinement is established, and this confinement can be
conveniently described by a scalar potential $m(r)$ that limits the
amplitude of the fermions to the region around the flux tube, as in
previous descriptions \cite{Wan88,Pav91,Gat92,Won93,Won95}. The Dirac equation for
a fermion in the tube in cylindrical coordinates $(r,\varphi,z)$ is:
\begin{eqnarray}
\label{eq:dir}
\left \{  \gamma^\mu  (\pi_\mu -eA_\mu) -m(r)
 \right \}  \Psi(r,\varphi,z,t)=0.
\end{eqnarray}
Following the results of Ref. \cite{Wan88,Won95}, we seek a solution of the
Dirac equation (\ref{eq:dir}) in the form
\begin{eqnarray}
\label{eq:psiphi}
\Psi(r,\varphi,z,t)=
\biggl [\gamma^\mu (\pi_\mu-eA_\mu)+m(r) \biggr ]\psi(r,\varphi,z,t).
\end{eqnarray}
The equation for $\psi$ is
\begin{eqnarray}
\label{eq:ephi}
\biggl \lbrace (p_0-eA_0)^2- (p_z-eA_z)^2
+i\alpha^3 e(\partial_z  A_0-\partial_0 A_z)
+ \bbox{p}_\perp^2-m^2(r)
+i \biggl [(\gamma^1\partial_1+\gamma^2 \partial_2)m(r)\biggr 
]
\biggr \rbrace  \psi(r,\varphi,z,t) = 0,
\end{eqnarray}
where $e$ is the coupling constant. We note that $ [\alpha^3, J_z] =
0,$ where $J_z = -i \partial / \partial\phi + \sigma_z/2 $ is the
third component of the angular momentum operator.  Furthermore, both
$J_z$ and $\alpha^3$ commute with the operator acting on
$\psi(r,\phi,z,t)$ in Eq.\ ({\ref{eq:ephi}).  Upon using the
  representation in Ref. \cite{Wan88}, the eigenfunction of $\alpha^3$
  satisfying $\alpha^3 \mu_\lambda=\eta_\lambda \mu_\lambda$ are
\begin{eqnarray}
\label{eq:rho}
  \mu_1  = \frac{1} {\sqrt{2}}
      \begin{pmatrix}  1       \cr
                       0       \cr
                       1       \cr
                       0       \cr
      \end{pmatrix},
{}~~\mu_2  = {1 \over \sqrt{2}}
             \begin{pmatrix}   0       \cr
                       1       \cr
                       0       \cr
                      -1       \cr       \end{pmatrix},
{}~~\mu_3  = {1 \over \sqrt{2}}
             \begin{pmatrix} 1       \cr
                       0       \cr
                      -1       \cr
                       0       \cr       \end{pmatrix},
{}~~\mu_4  = {1 \over \sqrt{2}}
             \begin{pmatrix} 0       \cr
                       1       \cr
                       0       \cr
                       1       \cr      \end{pmatrix},
\end{eqnarray}
with $\eta_{1,2}=+1$ and
with $\eta_{3,4}=-1$.
Therefore, we may choose $\psi(r,\varphi,z,t)$ to be
factorized as
\begin{eqnarray}
\label{eq:psi2}
\psi_{J_z}(r,\varphi,z,t)=
\sum_{\eta=-1,1} f_{J_z\eta}(z,t) R_{J_z\eta}(r,\varphi),
\end{eqnarray}
with $R_{J_z \eta}(r,\varphi)$ to be simultaneous eigenfunctions of
$J_z$ and $\alpha^3$.  The eigenfunctions of $J_z$ satisfying $J_z
R_{J_z \eta}=(\nu+\sigma_z/2) R_{J_z \eta}$ are
\begin{mathletters}
\begin{eqnarray}
\label{r1}
R_{J_z ~1}(r,\phi) = g_{1\nu}(r) e^{i\nu\phi} \mu_1
              -g_{2\nu}(r) e^{i(\nu+1)\phi} \mu_2 ,
\end{eqnarray}
\begin{eqnarray}
\label{r2}
R_{J_z -1} = g_{1\nu}(r) e^{i\nu\phi} \mu_3
              +g_{2\nu}(r) e^{i(\nu+1)\phi} \mu_4. 
\end{eqnarray}
\end{mathletters}

As a result of the transverse confinement, the gauge fields $A_0$ and
$A_z$ are confined within the transverse dimensions of the flux tube.
For high-energy collisions, the transverse dimensions of the flux tube
are much smaller than the longitudinal dimension.  To study the
dynamics along the longitudinal direction, it is reasonable to average
the gauge fields $A_0$ and $A_z$ over the transverse profile of the
flux tube.  After such a transverse averaging, the dynamics of $A_0$
and $A_z$ along the longitudinal direction can be approximated to be
independent of the transverse coordinates.  One can then use the
method of the separation of variables to separate the equation of
motion.  By introducing the transverse eigenvalue $m_\perp$, the Dirac
equation can be separated into the set of equations in different
coordinates,
\begin{eqnarray}
\label{ez1}
[p_0- A_0(z,t)]^2 -[p_z-e A_z(z,t)]^2 -m_{\perp}^2  -\eta  i
e[\partial_z A_0(z,t)-\partial_0 A_z(z,t)] f_{J_z\eta}(z,t)=0,
\end{eqnarray}
\begin{eqnarray}
\label{tran1}
\biggl [\bbox{p}_\perp^2(\nu) + m^2(r) - m_{\perp}^2 \biggr ]g_{1\nu}(r)
       = i{{\partial m(r)}\over{\partial r}} g_{2\nu} (r) ,
\end{eqnarray}
\begin{eqnarray}
\label{tran2}
\biggl [\bbox{p}_\perp^2(\nu+1) + m^2(r) - {m_{\perp}}^2\biggr ]g_{2\nu}(r)
       =- i{{\partial m(r)}\over{\partial r}} g_{1\nu}(r) ,
\end{eqnarray}
where
\begin{eqnarray}
\bbox{p}_\perp^2(\nu)
= -{{1}\over{r}}
{{\partial}\over{\partial r}}(r{{\partial}\over{\partial r}})
                        + {{\nu^2}\over{r^2}} \,.
\end{eqnarray}
Here, $m_\perp$ is the eigenvalue for the coupled transverse equations
(\ref{tran1}), and (\ref{tran2}), obtained by imposing the boundary
condition that the transverse wave functions $g_{1\nu}$ and $g_{2\nu}$
are transversely confined with a vanishing probability at $r\to
\infty$.  The eigenvalue $m_\perp$ depends on $J_z$ and is independent
of the quantum number $\eta$.  Some examples of $m(r)$, $m_\perp$ and
transverse wave functions have been presented previously
\cite{Gat92,Won93,Won95}.
 
We can write the wave function $\psi$ with the quantum number $J_z$
and a mass $m_\perp$ as a two-component wave function in an
abstract two-dimensional QED2 space as
\begin{eqnarray}
\psi_{\rm qed2}=\left ( \begin{matrix} f_{J_z\, 1}\cr f_{J_z\, -1}\cr 
\end{matrix}\right ) . 
 \end{eqnarray} 
In terms of this wave function, Eq.\ (\ref{ez1}) becomes
\begin{eqnarray}
\label{qed1}
\biggl \lbrack \gamma_{\rm qed2}^0 [p_0 - e A_0(x^1,t)] + \gamma_{\rm
  qed2}^1 [p_1 - e A_1(x^1,t)] - m_\perp \biggr \rbrack \psi_{\rm
  qed2}(x^1,t)=0,
\end{eqnarray}
where we re-label the longitudinal $z$-axis as the $x^1$-axis in QED2, and
\begin{eqnarray}\gamma_{\rm qed2}^0=\left(\begin{matrix} 0&1\cr
                1&0\cr \end{matrix}\right), \end{eqnarray}
\begin{eqnarray}\gamma_{\rm qed2}^1=i\sigma_2=\left(\begin{matrix} 0&1\cr
                -1&0\cr \end{matrix}\right)\,. \end{eqnarray} For
brevity of notation, the subscript `qed2' will be omitted.  It should
be kept in mind that the transverse state with $J_z$ in different
transverse excitations correspond to QED2 with different $m_\perp$.
We shall be interested in the state with the lowest $m_\perp$.

The above discussions shows how the fermion and the gauge field in
QCD4 in a flux tube can be approximately mapped into elements in QED2
for high-energy processes.  Although the non-Abelian nature of the
gauge field in QCD is needed to lead to the formation of the confining
flux tube, the non-Abelian property is not needed for particle
production in QED2 at high-energies.  An Abelian QED2 field theory possesses
the desirable properties of confinement and charge screening, and it
suffices to describe the particle production process at high energies.
Furthermore, in the non-Abelian field tensor
\begin{eqnarray}
F_{01}^i=\partial_0 A_1^i -\partial_1 A_0^i 
+g f^{ijk} A_0^j A_1^k, 
\end{eqnarray}
the non-linear quadratic term contains the product of $ A_0^j$ and
$A_1^k$.  One can convenient choose the Coulomb gauge $A_1^k=0$ such
that the non-linear quadratic term does not contribute.
We can therefore ignore the non-Abelian nature of the gauge fields and
approximate them to be
\begin{eqnarray}
\label{qed2}
F_{\mu\nu}=\partial_\mu A_\nu -\partial_\nu A_\mu ,
\end{eqnarray}
where $\mu,\nu=0,1$.  
The fermions give rise to
the current
\begin{eqnarray}
j^\mu = e \bar \psi \gamma^\mu \psi,
\end{eqnarray} 
which generates the gauge fields according to
\begin{eqnarray}
\label{qed3}
\partial_\nu F^{\mu \nu} =-j^\mu. 
\end{eqnarray}
Eqs.\ (\ref{qed1}), (\ref{qed2}), and (\ref{qed3}) constitute the
equations for the quantum mechanical system of QED2 with a fermion of
mass $m_\perp$. Thus, by assuming QCD confined within a flux tube, the
longitudinal dynamics of the system can be approximated as those of
QED2 with a mass $m_\perp$.  The gauge fields $A^\mu$
($\mu=0,1$) depend on the fermion field $\psi$. The fermion field
$\psi$, in turn, depends on the gauge field $A^\mu$. The coupling is
quite complicated and leads to a non-linear problem of great
complexity.  Remarkably, Schwinger found that QED2 involving massless
fermions with the gauge interaction is equivalent to a free boson
field $\phi$ with a mass $\mu_0=e/\sqrt{\pi}$, where $e$ is the
coupling constant \cite{Sch62}.

\section{\bf  Particle Production as an Initial-Value Problem in
Bosonized QED2}

In mapping elements of QCD4 approximately into elements of massive
QED2, what is the relationship between the coupling constant $g$ in
QCD4 and the coupling constant $e$ in QED2?  By limiting the motion and
the source distribution to reside in the longitudinal direction, the
coupling constant $e$ in QED2 acquire the dimension of a mass. The
confinement property is a non-perturbative property of QCD4.  The
coupling constant $e$ in QED2 should therefore be non-perturbatively
related to $g$.  The relationship can be retrieved by comparing
non-perturbative quantities.  In QED2 with the Coulomb gauge $A_1=0$,
the interaction energy between a quark and an anti-quark separated at
a separation of $x$ is $e^2 A_0(x)=e^2 x/2$.  On the other hand, in
QCD4, the non-perturbative confining interaction energy between a
quark and an anti-quark is $b x$ where $b$ is the string tension.
Therefore, equating the two interaction energies, we find a relation
between $e$ in QED2 and the non-perturbative string tension $b$ in
QCD4,
\begin{eqnarray}
e=\sqrt{2b}.
\end{eqnarray}
If we take the string tension to be $b=1$ GeV/fm, then $e=0.628$ GeV.
The boson mass in massless QED2 is $\mu_0=e/\sqrt{\pi}=0.354$ GeV.

The case of massive QED2 with a fermion mass $m_\perp$ can be studied
by bonsonization.  It is equivalent to the system of free bosons of
mass $\mu_0$ interacting with an interaction that depends on $m_\perp$
\cite{Col75,Col76}.  Does QCD corresponds to the case of strong
coupling with $e \gg m_\perp $, or the case weak coupling with $e \ll
m_\perp $?  The case of strong coupling is characterized by a
quasi-free bosons with confining fermions and charge screening, while
the limit of weak coupling approaches free Dirac theory with almost
free fermions dressed up as bosons having a mass close to the free
fermion rest mass \cite{Col76}.

We can estimate $m_\perp$ to be of the order of $\hbar/$(tube radius)
where the flux tube radius is of order 1 fm, leading to $m_\perp \sim$
0.2 GeV.  We have $\mu_0 \gg m_\perp$ which correspond to the case of
strong coupling with fermion confinement and color-charge screening,
rather than quasi-free Dirac particles.  Accordingly, the
mass-perturbation theory can be used to discuss the particle
production process in our case of massive QED2.

In the mass-perturbation theory, the unperturbed theory is massless
QED2 and the mass $m_\perp$ is treated as a perturbation.  Up to the
second order in $m_\perp$, the mass perturbation theory gives a
quasi-free boson with a mass $M$ given by \cite{Ada03}
\begin{eqnarray}
M^2=\mu_0^2 + 2 e^{\gamma} \mu_0 m_\perp + 1.0678e^{2 \gamma} m_\perp^2,
\end{eqnarray}
where $\gamma=0.5772$ is the Euler constant. We therefore have $M\sim
\mu_0+ e^{\gamma} m_\perp$.  For our case of $\mu_0=0.354$ GeV and
$m_\perp\sim 0.2$ GeV, we get $M\sim 0.71$ GeV, which comes close to
the spin-spin averaged mass of 0.62 GeV for the $\pi$-$\rho$ pair.
Thus, the boson of massive QED2 finds its correspondence as the boson
in QCD that splits into $\pi$ and $\rho$ when the spin-spin
interaction is taken into account.

As mass perturbation theory is based on massless QED2 with $m_\perp$
as a perturbation, the application of the theory to particle
production process involves in using the results of massless QED2 and
replacing the boson mass $\mu_0$ in these massless QED2 results by the
corrected mass $M$.  In practical applications, this amounts to
replacing $\mu_0$ with the physical mass, including the effects of the
effective mass increase due to the transverse momentum.  As pions are
the most predominantly produced particle, the phenomenological
treatment then involves in replacing $\mu_0$ by
$M=\sqrt{m_\pi^2+p_{\perp,\pi}^2}$.

We can review the rapidity distribution for massless QED2 obtained
previously \cite{Won95}.  The relation between the bosonic and the
fermionic quantities in massless QED2 is \cite{Cas74,Col75,Col76}
\begin{eqnarray}
j^{\mu}=-e\epsilon^{\mu \nu} \partial_{\nu}\phi/\sqrt{\pi},
\end{eqnarray}
where $j^{\mu}$ is the fermionic current which can be taken to be a
real quantity, and $\epsilon^{\mu \nu}$ is the antisymmetric tensor
$\epsilon^{01}= -\epsilon_{01}=-1 $.  We note that, as $j^\mu$ is a
vector field and $\epsilon^{\mu \nu}$ is a pseudotensor, the field
$\phi$ is a real pseudoscalar field, and it represents the color
electric field $F^{01}$, as $F^{01}=e\phi/\sqrt{\pi}$.  If the current
$j^\mu$ arising from the fermions is initially known, then the
dynamics of the pseudoscalar field $\phi$ can be inferred at all
times.  Treating the problem as a system of quasi-free bosons with a
mass $M$, the initial value conditions will allow us to determine the
dynamics of the system.  To apply the results to our case, we will
work within mass perturbation theory which is a quasi-free boson
system with $e/\sqrt{\pi}$ replaced by $M$.

Given an initial fermion charge distribution $j_\mu(x,t=0)$, its
Fourier transform is $\tilde {j^\mu}(p^1)$ is
\begin{eqnarray} 
\tilde {j^\mu}(p^1) = {1 \over \sqrt{2\pi}} \int dx e^{-ip^1x} { j^\mu(x,0) }.
\end{eqnarray}
We show previously \cite{Won91} that the momentum distribution of the bosons is
then given by 
\begin{eqnarray}{dN \over dp^1}={\pi \over 2p^0e^2} 
\biggl \lbrack {p^0 \over p^1} \tilde {j^0} (p^1) + \tilde {j^1}(p^1)  
\biggr \rbrack
\biggl \lbrack {p^0 \over p^1} \tilde {j^0} (-p^1) + \tilde {j^1}(-p^1)  
\biggr \rbrack,
\end{eqnarray}
and the rapidity distribution of the produced particles is
\begin{eqnarray}{dN \over dy}=
{\pi \over 2e^2} 
\biggl \lbrack {p^0 \over p^1} \tilde {j^0} (p^1) + \tilde {j^1}(p^1)  
\biggr \rbrack
\biggl \lbrack {p^0 \over p^1} \tilde {j^0} (-p^1) + \tilde {j^1}(-p^1)  
\biggr \rbrack.
\end{eqnarray}
This gives a simple relation between the rapidity distribution and the
Fourier transforms of the initial fermionic charge current.

We can review how this initial-value problem in massless QED2 can be
formulated for the case of a positive charge $\nu e$ separating from a
negative charge $-\nu e$ with a center-of-mass energy $\sqrt {s}$
\cite{Won91}.  We work in the center-of-mass system and start at $t=0$
with the charge and anti-charge pair superimposed so that the total
charge density of the system at $t=0$ is zero:
\begin{eqnarray}j^0(x,0)=0. 
\end{eqnarray}
To construct the initial longitudinal current, we introduce a
distribution that depend on $\sigma$.
\begin{eqnarray}
j^1(x,t)=
\frac{\partial}{\partial t}
\left \{ {\nu\over2}[\tanh ( (x + t)/\sigma)+1]
+ {\nu\over2}[\tanh ( (x - t)/\sigma)+1]
\right \}.
\end{eqnarray}
In this case, the initial current which arises from a charge $\nu e$
moving in the positive $x$ direction and another charge $-\nu e$
moving in the negative $x$ direction is given by
\begin{eqnarray}j^1(x,0)= {\nu e\over {\sigma \cosh ^2 ({x/\sigma}}) }. 
\end{eqnarray}
In the limit as $\sigma$ approaches zero, the above current is
proportional to a delta function.  The diffusivity $\sigma$ is related
to the total invariant mass $\sqrt {s}$ of the system; using the
energy $P^0=\sqrt s$ at the initial time $t=0$, we obtain a relation
between $\sigma$ and $\sqrt s$:
\begin{eqnarray}
\sigma={ 2\pi \nu^2 \over {3\sqrt{s}}}. 
\end{eqnarray}
For this current 
distribution $j^\mu(x,0)$ the Fourier transform of $j^1(x,0)$ is 
\begin{eqnarray}
\tilde {j^1} (p^1)
={ { \nu e\pi p^1 \sigma} \over { \sqrt{2\pi} \sinh (\pi p^1 \sigma /2)} },
\end{eqnarray}
and the rapidity distribution in massless QED2 is \cite{Won91}
\begin{eqnarray}
\label{dndy}
{ dN \over dy } ={ {\nu^2 \xi^2} \over \sinh^2{\xi} }, 
\end{eqnarray}
where
\begin{eqnarray}
\label{mass}
\xi={ { \nu^2 \pi^2 \mu_0 \sinh y} \over {3 \sqrt{s}} }. 
\end{eqnarray}
The rapidity distribution therefore shows a plateau structure around
$y\sim 0$.  In the limit of very high energy, the rapidity
distribution is $dN/dy=\nu^2$, which agrees with the result of Casher
\cite{Cas74} (for $\nu=1$).  

Within the mass perturbation theory, we can approximate the particle
production process of massive QED2 using the results from massless
QED2 and replacing the $\mu_0$ of massless QED2 in Eq.\
(\ref{mass}) by $M$ in massive QED2, with the result
\begin{eqnarray}
\label{mass1}
\xi={ { \nu^2 \pi^2 M \sinh y} \over {3 \sqrt{s}} }. 
\end{eqnarray}
Using Eqs.\ (\ref{dndy}) and (\ref{mass1}) by replacing $\mu_0$ with
the phenomenological mass $M=\sqrt{m_\pi^2+\langle
p_{\perp,\pi}\rangle^2}=0.30$ GeV and $\nu=2.45$ gives a good
phenomenological fit to the $dN_{\rm \pi^\pm}/dy$ data in $e^+$-$e^-$
experiment at $\sqrt{s}=29$ GeV (in Fig.\ (40) of \cite{Aih88}).

\section{Evolution of the Medium Parton Momentum Distribution}

We conclude from our discussions in the last two sections that in
addition to experimental evidences for the rapidity plateau in
elementary QCD particle production processes, theoretical
investigations in plausible models also show the occurrence of a
rapidity plateau when a color charge pulls away from an anti-color
charge at high energies.  As a nucleus-nucleus consists of elementary
production processes of string fragmentation, there can be a similar
plateau structure in the rapidity distribution of the produced medium
partons, consistent with the parton rapidity plateau we have extracted
at the early stage of the nucleus-nucleus collision.

In nucleus-nucleus collisions, this early parton momentum distribution
can be probed by a jet produced in the early stage of the collision.
Those medium partons kicked by the jet subsequently materialize as
ridge particles and they retain the property of the rapidity plateau. 

The plateau rapidity structure of the early parton momentum
distribution differ from the Gaussian rapidity distribution of the
bulk matter \cite{Mur04,Ste05,Ste07}.  How does one understand such a
difference?

It is important to point out that the early momentum distribution
represents the momentum distribution at the early stage of the
nucleus-nucleus collision as it involves the direct reaction with the
jet, which occurs only at the early stage of the nucleus-nucleus
collision.  On the other hand, the momentum distribution of the bulk
matter represent the momentum distribution of the bulk matter at the
end-point of the nucleus-nucleus collision.  A considerable period of
time separates the beginning, early stage of the nucleus-nucleus
collision and the end-point of the nucleus-nucleus collision.
Significant dynamical evolution must have occurred between these two
separate time points, as described schematically in Fig. 9 of
Ref. \cite{Won07}.  The time evolution of the momentum distribution
will make the end-point momentum distribution of the bulk matter
different from the early parton momentum distribution.

Evidence for the occurrence of a dynamical evolution of the momentum
distribution presents itself in the difference of (i) the transverse
momentum distribution extracted at the moment of the jet-(medium
parton) collisions, and (ii) the transverse momentum distribution of
the bulk matter at the end-point of the nucleus-nucleus collision.
The extracted early parton transverse momentum distribution, as given
by Eqs.\ (\ref{dis2}), and (\ref{bestset1}), has a thermal-like
distributions, with an initial inverse slope $T=0.5$ GeV that is
slightly greater than the inverse slope of the end-point transverse
momentum distribution, consistent with the direction of transverse
momentum evolution from a higher inverse slope $T$ to a lower inverse
slope $T$ value \cite{Put07}.  We expect that the rapidity
distribution will likewise evolve and its shape will change with time.
There is no reason to expect that the longitudinal momentum
distribution at the early stage of the nucleus-nucleus collision
should be the same as the corresponding longitudinal momentum
distribution at the end-point of the nucleus-nucleus collision.

To understand the evolution of the medium parton momentum
distributions, we should think of the full momentum distribution to be
a six-dimension distribution function $F({\bf r},{\bf p},t)$ of the
medium that depends both on the spatial and momentum coordinates, as
well as on the lapsed time $t$.  The parton momentum distribution
extracted here is in effect an average of this the six-dimensional
distribution function $F$ over spatial and temporal coordinates of the
collision points in the early stage of the nucleus-nucleus collision,
using the jet as a probe.  After the early stage of jet-(medium
parton) collisions, partons from one position will collide with
partons of adjacent positions.  These collisions will modify the
momenta of the colliding partons, leading to a change of the
distribution function $F({\bf r},{\bf p},t)$ as a function of time.
How the evolution will take place is a problem of great complexity
that depends on models with many unknown theoretical elements
\cite{Mod08}.  Nevertheless, one expects that starting with a
non-isotropic plateau rapidity distribution that is much elongated in
the longitudinal direction, a collision of two partons with large and
opposing longitudinal momenta in adjacent spatial locations will
redistribute the partons from the longitudinal direction towards the
transverse directions, with a decrease in the longitudinal momenta of
the colliding partons.  Hence, the evolution will smooth out the
anisotropic plateau rapidity structure to a significant degree as time
proceeds.

\section{Dependence of the Fragmentation Function on 
the Jet-(medium parton) Collision Number}

We turn now to investigate the geometry-dependent part of the ridge
and trigger particle yields, as given previously in Eqs.\
(\ref{trig}), (\ref{avern}), and (\ref{eq11}).  The jet fragmentation
function in these equations depends on the number of collisions $N$
(or $N_k$) suffered by the jet parton, and the observed physical
quantities depend on the distribution $P(N)$.  We envisage jet partons
to be produced by binary nucleon-nucleon hard-scattering processes and
we focus our attention on one of the jet partons.  We consider the jet
parton to occur at $\eta_{\rm jet}=0$ such that the jet momentum ${\bf
p}_j$, the trigger particle momentum ${\bf p}_{\rm trig}$, and the
momentum kick ${\bf q}$ all lie in the mid-rapidity transverse plane
in the collider system, pointing in the same direction.  The vector
symbol for these quantities can be understood.

We envisage that in the passage of the parent jet parton in the dense
medium, the jet parton with initial momentum $p_j$ imparts a momentum
$q$ to each kicked medium parton, and loses a momentum $Nq$ after $N$
jet-(medium parton) collisions.  In addition to collisional momentum
loss, the jet parton can lose momentum by gluon radiation
\cite{Jetxxx}.  As the radiated gluon will likely come out in a cone
along the jet direction in random azimuthal angles, the average
momentum loss due to gluon radiation $\langle {\bf \Delta}_r\rangle$
lies along the jet direction ${\bf e}_j$.  We can parametrize the
radiative gluon momentum loss phenomenologically by $|{\bf
\Delta_r}|=Nq_r$ where the $q_r$ value obtained from experimental data
will need to be compared with theoretical models.  In practice, the
collisional and radiative momentum losses appear together as the sum
total $(q+q_r)$ in the fragmentation function [Eq.\ (\ref{qqr})].
Furthermore, there can be additional attenuation $\zeta_a$ due to
absorptive inelastic processes of removing the jet from the jet 
channel.  Only the sum of the absorptive, collisional, and radiative
contributions, leading to the total attenuation coefficient $\zeta$,
can be obtained by comparison with experimental jet quenching data
[Eqs.\ (\ref{zetasum}) and (\ref{nzeta})].

Upon including the momentum loss due to jet-(medium parton) collisions and
gluon radiation in the momentum kick model, Eq. (\ref{trig}) becomes
\begin{eqnarray}
\label{trig1}
N_{\rm trig} \!\!= \!\!\int \! d{\bf p}_j \frac {dN_j}{d{\bf p}_j}
\!\sum_{N\!=\!0}^{N_{\rm max}} \!P(N) e^{-\zeta_a N} D({ p}_{\rm
trig};{ p}_j \!-\! N ({ q} +{ q}_{r}) ).
\end{eqnarray}
We wish to write out the dependence of the fragmentation function $D$
on $N$ in the above equation explicitly.  The dominant contribution of
the jet production process comes from gluon-gluon collisions
\cite{Owe78}.  The relevant fragmentation function of fragmenting a
pion out of a gluon at the momentum scale $Q_0^2$, can be written in
the form \cite{Owe78}
\begin{eqnarray}
z D(z,Q_0^2)\sim C_\pi (1-z)^{a_1},
\end{eqnarray}
where $z=p_{\rm trig}/p_j$ and $C_\pi$ is a constant.  In perturbative
QCD, the fragmentation function near $z=1$ varies with the QCD momentum
scale $Q$ according to \cite{Owe78,Gro74}
\begin{eqnarray}
zD(z,Q^2)\simeq
zD(z,Q_0^2)e^{0.69G{\bar s}}(-\ln z)^{4G\bar s}
\frac {\Gamma (a_1 + 1)} {\Gamma{(a_1+1+4G\bar s)}},
\end{eqnarray}
where $G=4/25$, and $\bar
s=\ln[\ln(Q^2/\Lambda^2)/\ln(Q_0^2/\Lambda^2) ]$.  After the jet
suffers $N$ jet-(medium parton) collisions with the medium partons,
the fragmentation function for a pion to fragment out of the final jet
of momentum $z_N=p_{\rm trig}/[p_j-N(q+q_r)]$ is
\begin{eqnarray}
D(z_N,Q^2)
=C_\pi e^{0.69G{\bar s}} \exp \{-\ln z_N+ a_1 \ln (1-z_N)
+{4G\bar s}\ln[(-\ln z_N)]\}
\frac {\Gamma (a_1 + 1)} {\Gamma{(a_1+1+4G\bar s)}}.
\end{eqnarray}
Upon expanding the exponent index of the above function in power of
$N(q+q_r)/p_j$ and retaining the term first order in $N(q+q_r)/p_j$,
we obtain the dependence of the jet fragmentation function on the
jet-(medium parton) collision number $N$,
\begin{eqnarray}
D(p_{\rm trig}/(p_j-N(q+q_r)),Q^2)\simeq
D(p_{\rm trig}/p_j,Q^2)e^{-\zeta_D N} ,
\end{eqnarray}
where 
\begin{eqnarray}
\label{qqr}
\zeta_D = \left ( \frac{1}{p_{\rm trig}/p_j} 
  + \frac {a_1}{1-{p_{\rm trig}/p_j}}
+\frac{ 4G\bar s}{\ln{(p_j/p_{\rm trig})}} \frac {p_j}{p_{\rm trig}}\right )
\frac {p_{\rm trig} (q+q_r)}{p_j^2}.
\end{eqnarray}
Substituting this relationship into Eq.\ (\ref{trig1}), we get
\begin{eqnarray}
N_{\rm trig} \!\!= \!\!\int \! d{\bf p}_j \frac {dN_j}{d{\bf p}_j} 
 D({\bf p}_{\rm trig};{\bf p}_j)
\!\sum_{N\!=\!0}^{N_{\rm max}}
\!P(N) 
e^{-\zeta N}, 
\end{eqnarray}
where we have combined $\zeta_a$ with $\zeta_D$,
\begin{eqnarray}
\label{zetasum}
\zeta=\zeta_a+\zeta_D.
\end{eqnarray}  
Because of the normalization condition Eq.\ (\ref{dnorm}) for the
fragmentation function and the definition of $dN_j/d{\bf p}_j$ in Eq.\
(\ref{Nj}), we obtain
\begin{eqnarray}
N_{\rm trig}=
N_{\rm bin}\!\sum_{N\!=\!0}^{N_{\rm max}}
\!P(N) 
e^{-\zeta N}, 
\end{eqnarray}
and we get the jet quenching measure
\begin{eqnarray}
\label{nzeta}
R_{AA}=\frac{N_{\rm trig}} {N_{\rm bin}}
=\sum_{N\!=\!0}^{N_{\rm max}}
\!P(N) 
e^{-\zeta N}, 
\end{eqnarray}
and the average number of jet-(medium parton) collisions per trigger
\begin{eqnarray}
\label{eq37}
 \langle N \rangle = \langle N_{k} \rangle =
\sum_{N\!=\!1}^{N_{\rm max}}
\! N P(N) 
e^{-\zeta N} \biggr /
\sum_{N\!=\!0}^{N_{\rm max}}
\!  P(N) 
e^{-\zeta N}.
\end{eqnarray}

The presence of the attenuation factor $e^{-\zeta N}$ implies that
detected trigger particles are likely to originate near the surface
where the number of jet-(medium parton) collisions $N$ is smallest.
The quantity $\zeta_a$ is not known.  The quantity $\zeta_D$ depends
on $q+q_r$ and $p_j$.  We can estimate the contribution of collisional
contribution to the value of $\zeta_D$ by inferring the approximate
average value of $p_j$ for $p_{\rm trig}\sim 5$ GeV.  As $p_j \sim
(p_{\rm trig} + \langle N \rangle q + 3N_{\rm jet} T_{\rm jet})$ with
$\langle N\rangle \sim 6$, $q\sim 1$ GeV, $N_{\rm jet}$=0.75, $T_{\rm
jet}=0.55$ GeV, we estimate that $p_j\sim 2.5 \, p_{\rm trig}$.  If we
use $Q^2=p_j^2$, $Q_0^2=3$ GeV$^2$, $a_1=1.5$ and $\Lambda=0.5$ GeV as
in \cite{Owe78} we can use Eq.\ (\ref{qqr}) and estimate the
contribution to $\zeta_D$ from $q$ to be approximately 0.22.  There
can be additional contributions from the radiative energy loss $q_r$.
We shall set $\zeta$ as a free parameter to describe the experimental
$R_{AA}$ data by searching for $\zeta$ around the neighborhood of
about 0.22.  We find in Section XIII that the experimental jet
quenching and ridge yield data are consistent with a value of
$\zeta=0.20$ which comes very close to the value of 0.22 estimated
here.

\section{Geometry Dependence of Trigger and Ridge Yields}

Because the ridge particle yield has been measured on the basis of the
yield per trigger particle, it is necessary to determine the trigger
yield $N_{\rm trig}$ as a function of the geometrical variables, in
addition to determining the number of ridge particles.  The
trigger particle yield is quenched due to the energy loss of the jet
parton. We therefore need to study jet quenching and follow the
trajectory of the jet.

From our earlier considerations, the relevant physical quantities are
given in terms of $N$, the number of medium partons kicked by the jet
on its way to emerge from the medium.  This quantity $N$, in turn, is
equal to the number of jet-(medium parton) collisions $N_k$ suffered
by the jet parton.  We assume for simplicity that the energetic jet
parton travels along a straight line trajectory with a velocity nearly
the speed of light, making an angle $\phi_s$ with respect to the
reaction plane, $\phi_s=\phi_{\rm jet}-\phi_{\rm RP}$.  Using the
mid-point $O$ between the centers of the two nuclei as the origin, we
set up a transverse coordinate system for the jet source point ${\bf
b}_0$ and the jet trajectory point ${\bf b}'$ as shown in Fig.\ 7.
\begin{figure} [h]
\includegraphics[angle=0,scale=0.60]{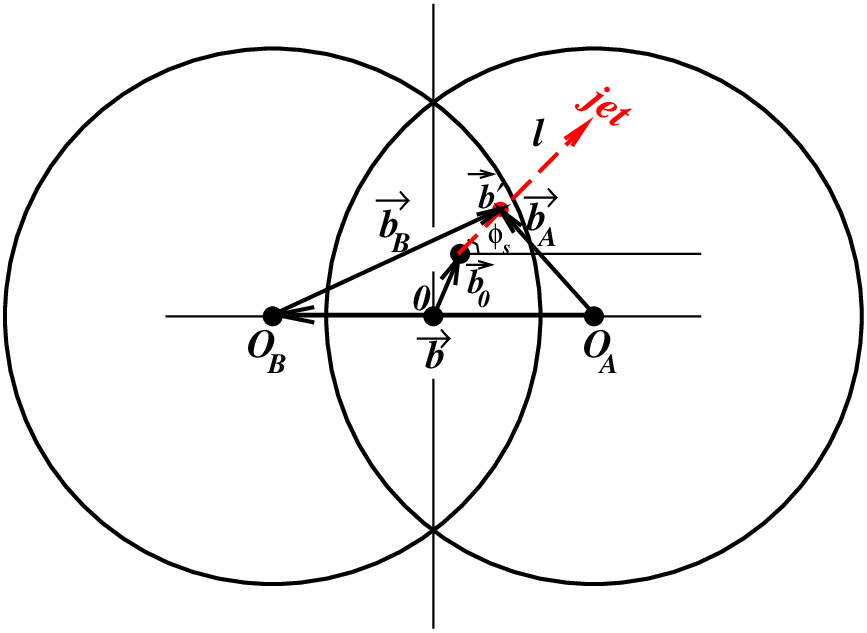}
\vspace*{0.0cm}
\caption{ (Color online) The transverse coordinate system used for the
jet source point ${\bf b}_0$ and the trajectory point ${\bf b}'$.  The
coordinate origin is located at the midpoint $O$ between the two
colliding nuclei, whose centers are located at $O_A$ and $O_B$
separated by an impact parameter ${\bf b}$ . The jet trajectory lies
along $l$ and makes an angle $\phi_s$ with respect to the reaction
plane. }
\end{figure}

We consider the jet source point at ${\bf b}_0$, from which a
mid-rapidity jet parton originates.  The number of jet-(medium parton)
collisions along the jet trajectory making an angle $\phi_s$ with
respect to the reaction plane is
\begin{eqnarray}
\label{nk}
N_k({\bf b}_0,\phi_s) = \int_0^{\infty} \sigma \, dl
\frac{dN_{\rm parton}}{dV} \left [{\bf b}' ({\bf b}_0,\phi_s) \right ],
\end{eqnarray} 
where $0<l<\infty$ parametrizes the jet trajectory, $\sigma$ is the
jet-(medium parton) scattering cross section, and $dN_{\rm
parton}({\bf b}')/{dV}$ is the parton density of the medium at ${\bf
b}'$ along the trajectory $l$.  Jet-(medium parton) collisions take
place along different parts of the trajectory at different $l$ and
involve the medium at different stages of the expansion.  They depend
on the space-time dynamics of the jet and the medium.  To follow the
jet-(medium parton) collisions along the jet trajectory, we need a
time clock to track the coordinates of the jet and the motion of the
medium.  We start the time clock for time measurement at the moment of
maximum overlap of the colliding nuclei, and the jet is produced by
nucleon-nucleon collisions at a time $t\sim \hbar/(10 {\rm GeV})$
which can be taken to be $ \sim 0$. The trajectory path length $l$ is
then a measure of the time coordinate, $t \approx l$, which is needed
to follow the longitudinal and transverse expansions of the medium.

The trajectory point ${\bf b}'$ depends on the jet source origin point
${\bf b}_0$ and the jet azimuthal angle $\phi_s$ as
\begin{eqnarray}
{\bf b}'({\bf b}_0,\phi_s) =(b_x', b_y') = (b_{0x}+l\cos
\phi_s,b_{0y}+l\sin \phi_s).
\end{eqnarray}

If we approximately represent the modification of the density arsing
from longitudinal and transverse expansion by an effective time
parameter $\tau_{\rm eff}$ using the initial parton density in the
following approximation,
\begin{eqnarray}
\label{nkk}
N_k({\bf b}_0,\phi_s) =\frac{\sigma}{\tau_{\rm eff}} \int_0^{\infty} dl
\frac{dN_{\rm parton}}{d{\bf b}'} \left [{\bf b}' ({\bf b}_0,\phi_s)]. \right .
\end{eqnarray} 
then we find that the data can be described by $\zeta=0.22$ and
$\sigma/\tau_{\rm eff} \sim 0.025$ fm.  Such a picture only gives a
rather crude description of the path-length dependence of the ridge
yield.

In order to give a more realistic picture, we describe the medium by
an expanding fluid with an initial density given by the distribution
of the participants at the moment of maximum nuclear overlap.  We
assume that the longitudinal expansion begins at the moment of maximum
overlap as the initial momenta are directed along the longitudinal
direction.  A period of time $t_0$ is however needed to convert the
longitudinal kinetic energy into entropy to produce particles with a
transverse mass. The transverse hydrodynamical expansion can then
commence at $t\ge t_0$.  The time for producing a particle with a
typical transverse mass of about 0.35 GeV is $\hbar/(0.35 {\rm
~GeV})\sim 0.6$ fm/c, which is also the time estimated for the
thermalization of the produced matter \cite{Hei02}.  We therefore take
$t_0=0.6 $ fm/c.

As we will focus our attention in the mid-rapidity region where
experimental data are available, Bjorken hydrodynamics \cite{Bjo83}
and Landau hydrodynamics \cite{Lan53,Bel56} coincide
\cite{Won08c,Won08d} and we can use Bjorken hydrodynamics to describe
the longitudinal expansion.  For a hydrodynamical system undergoing
Bjorken longitudinal expansion, the transverse expansion can be
described by the hydrodynamical solution of Baym $et~al.$
\cite{Bay83}.  Using the method of characteristics, they find that the
energy density and velocity field in the transverse direction can be
described well approximately by analytic formulas.

Accordingly, we follow the jet along its trajectory at $l$ specified
by the trajectory point ${\bf b}'$ with a transverse magnitude
$b'=|{\bf b}'|$ (measured from the origin $O$) at at time $t=l$.  The
time after the onset of the transverse expansion is then $t_R=t-t_0$,
and a rarefaction wave travels from the transverse radius $R$ inward
with the speed of sound $c_s$.  The dynamics is different whether the
rarefaction wave has reached this medium point $b'$ or not. The
transverse space of the medium can be divided into Region I and II.

In Region I characterized by $b' < R-c_s t_R$, the rarefaction wave
has not reached this medium point at ${\bf b}'$. In this region, the
medium has not started to expand transversely with transverse velocity
$v_\perp=0$ while the longitudinal expansion has already commenced.
Due to the longitudinal expansion the density is depleted and the
temperature is decreased as \cite{Bay83}
\begin{eqnarray}
T \propto (t_0/t)^{c_s^2}.
\end{eqnarray}
As the entropy density and number and entropy densities are
proportional to $T^{1/c_s^2}$, we have
\begin{eqnarray}
\label{reg1}
\frac {dN_{\rm parton}}{ dV}(b',t)
=\frac{ dN_{\rm parton}}{ dV} (b_{\rm init}',t=t_0) \frac {t_0}{t},
\end{eqnarray}
with $b_{\rm init}'=b'$ in Region I.

In Region II defined by $R-c_s t_R < b' < R+t_R$, where the
inward-traveling rarefaction wave has passed through already.  The
medium is expanding transversely outward and the transverse velocity
$v_\perp$ at the point ${\bf b}'$ at the time $t$ is
\begin{eqnarray}
v_\perp=\frac{b'-R+c_s t_R}{t_R+c_s(b'-R)}.
\end{eqnarray}
The transverse velocity $v_\perp$ is unity (speed of light) at the
surface point $R+t_R$, and is zero at at the point $b'-c_s t_R$ where
the rarefaction wave has just arrived.  The medium temperature in this
region is given by \cite{Bay83}
\begin{eqnarray}
T(b',t) =T_0(b_{\rm init}',t=t_0) 
\left ( \frac{t_R-b'+R}{t_R+b'-R} ~~\frac{1-c_s}{1+c_s} \right )^{c_s/2}
\left ( \frac{t_0}{t} \right ) ^d ,
\end{eqnarray}
where $b_{\rm init}'=b'-v_\perp t_R$ is the initial position that reaches
$b'$ at $t_R$, and if $b'-v_\perp t_R \le 0$ we set $b_{\rm init}'=0$. 
Here the exponential index $d$ is \cite{Bay83}
\begin{eqnarray}
d=\frac{c_s^2}{2}\left [1+\frac{1}{1-v_\perp(b',t) c_s}\right ].
\end{eqnarray}
The corresponding medium number density in the coordinate frame in
which the trajectory $l$ is measured (Fig.\ 7) is therefore
\begin{eqnarray}
\label{reg2}
\frac {dN_{\rm parton}}{dV} (b',t) =  \gamma
\frac {dN_{\rm parton}}{dV} (b_{\rm init}',t=t_0) 
\left ( \frac{t_R-b'+R}{t_R+b'-R}~~ \frac{1-c_s}{1+c_s} \right )^{1/2c_s}
\left ( \frac{t_0}{t} \right )^{d/c_s^2} ,
\end{eqnarray}
where 
\begin{eqnarray}
\gamma=\frac{1}{\sqrt{1-v_\perp^2}},
\end{eqnarray}
and $\gamma$ is to take into account the change in the medium number
density due to the flow velocity of the medium along the transverse
direction.  While the boundary $R$ is independent of the azimuthal
angle of ${\bf b}'$ for the central collision, the boundary radius is
a function of the azimuthal angle for non-central collisions.  We
shall assume that the relations between the density and the radius
given above remain applicable by using a radius $R$ that depends on
the azimuthal angle.  These results of the number density at various
transverse points allow one to obtain the absorption exponent index
for a jet to pass through an expanding medium.  In numerical
calculations, we take the speed of sound to be $c_s=1/\sqrt{3}$.

The medium parton density $dN_{\rm parton}/dV$ at $(b_{\rm
init}',t=t_0)$ is related to the parton transverse density $dN_{\rm
parton}/d{\bf b}$ at $t_0$ by
\begin{eqnarray} 
\frac {dN_{\rm parton}}{dV} (b_{\rm init}',t=t_0)
=\frac {dN_{\rm parton}}{2 t_0d{\bf b}'} (b_{\rm init}',t=t_0) 
\end{eqnarray}
We can relate the initial parton number transverse density ${dN_{\rm
parton}}/{d{\bf b}'}$ at $t=t_0$ to the corresponding participant
initial number transverse density ${dN_{\rm part}}/{d{\bf b}'}$ as
\begin{eqnarray} 
\label{eq40}
\frac{dN_{\rm parton}}{d{\bf b}'} = \frac{dN_{\rm parton}}{dN_{\rm
part}} \frac{dN_{\rm part}}{d{\bf b}'} =\kappa \frac{dN_{\rm
part}}{d{\bf b}'} ,
\end{eqnarray} 
where $\kappa$=$dN_{\rm parton}/dN_{\rm part}$ is the number of
partons per participant.  A previous collection of data gives $N_{\rm
ch}/\langle N_{\rm part}/2\rangle =28$ for $\sqrt{s_{NN}}=200$ GeV and
16 for $\sqrt{s_{NN}}=62$ GeV (see \cite{Bus04,Pho04}).  If we use the
parton-hadron duality and count the parton number by (3/2) times the
charged multiplicity of detected hadrons, then we get
\begin{eqnarray} 
\kappa=
\begin{cases}
21 & {\rm ~for~}\sqrt{s_{NN}}=200~{\rm GeV}\\
12 & {\rm ~for~}\sqrt{s_{NN}}=62~{\rm GeV}.\\
\end{cases}
\end{eqnarray} 

A given source point ${\bf b}_0$ and a given azimuthal angle $\phi_s$
will lead to $N_k({\bf b}_0,\phi_s)$ number of kicked medium partons,
which we shall identify by parton-hadron duality as ridge particles.
The jet number transverse density is given by the binary
nucleon-nucleon collision number transverse density, as
nucleon-nucleon collisions are the source of jets.  We need to weight
the number of kicked medium particles by the local binary collision
number element $d{\bf b}_0 \times dN_{\rm bin}/d{\bf b}_0$.  The
normalized probability distribution $P(N,\phi_s)$ with respect to the
number of ridge particles (or jet-(medium parton) collisions) is
\begin{eqnarray}
\label{Pn}
P \left( N,\phi_s \right ) 
=\frac{1}{N_{\rm bin}} \int d {\bf b}_0 \frac{dN_{\rm bin}}{d {\bf b}_0}
( {\bf b}_0) \delta [N-N_k({\bf b}_0,\phi_s)],
\end{eqnarray}
which leads to the desired normalization
of the distribution $P(N,\phi_s)$,
\begin{eqnarray}
\label{norm}
\int {dN} P(N,\phi_s)=1,
\end{eqnarray}
and the total number of binary nucleon-nucleon collisions
\begin{eqnarray}
N_{\rm bin}= \int d {\bf b}_0 \frac{dN_{\rm bin}}{d {\bf b}_0}.
\end{eqnarray}
Thus, the number of ridge particle yield per trigger particle (or the
number of jet-(medium parton) collisions per trigger) at an azimuthal
angle $\phi_s$, averaged over all source points of binary collisions
at all ${\bf b}_0$ points, is
\begin{eqnarray}
\label{eq44}
\bar N_{k} (\phi_s) = \int N P (N, \phi_s) 
e^{-\zeta N}
~dN \biggr / \int P (N, \phi_s) 
e^{-\zeta N}
~dN.
\end{eqnarray}
In practical calculations, it is convenient to discretize $N$ by
replacing the delta function in Eq.\ (\ref{Pn}) as
\begin{eqnarray}
 \delta (N-N_k)
\to \{ \Theta[N_k - (N-\Delta N/2)] - \Theta[N_k - (N+\Delta N/2)]\}/\Delta N. 
\end{eqnarray}
Upon choosing $\Delta N=1$, the normalization condition (\ref{norm})
becomes
\begin{eqnarray}
\sum_{N=0}^{N_{\rm max}} P(N,\phi_s)=1,
\end{eqnarray}
as in our previous definition with the $\phi_s$ dependence now
explicitly written out.  Equation (\ref{Pn}) in the discretized form
of $N$ becomes
\begin{eqnarray}
\label{Pna}
P \left( N,\phi_s \right ) 
=\frac{1}{N_{\rm bin}} \int d {\bf b}_0 \frac{dN_{\rm bin}}{d {\bf b}_0}
( {\bf b}_0) \{\Theta [(N-\Delta N/2)-N_k({\bf b}_0,\phi_s)]
-\Theta [(N+\Delta N/2)-N_k({\bf b}_0,\phi_s)]\}/{\Delta N}.
\end{eqnarray}
This equation facilitates the evaluation of $P(N,\phi_s)$.  For a
given $\phi_s$, we evaluate $N_k({\bf b}_0,\phi_s)$ at different
source points ${\bf b}_0$, place the quantity $dN_{\rm bin}/{d {\bf
b}_0}$ at the appropriate $[N-\Delta N/2 \le N_k({\bf b}_0,\phi_s) \le
N +\Delta N/2]$ bin, and accumulate the contributions from all jet
source points at all ${\bf b}_0$. The accumulated distribution,
divided by $N_{\rm bin} \Delta N$ is then the distribution function
$P( N,\phi_s)$.  For these calculations, we need the transverse
densities of the binary nucleon-nucleon collisions and participant
numbers.  The transverse density of binary nucleon-nucleon collisions
in Eq.\ (\ref{Pna}) can be obtained from the Glauber model to be
\begin{eqnarray} 
\frac{dN_{\rm bin}}{d{\bf b}_0} ({\bf b}_0)=ABT({\bf b}_{A0})T({\bf
b}_{B0}) \sigma_{\rm in}^{NN} ,
\end{eqnarray}
where ${\bf b}_{A0}={\bf b}_0+{\bf b}/2$ and ${\bf b}_{B0}={\bf
b}_0-{\bf b}/2$.  The quantity $\sigma_{\rm
in}^{NN}$ is the nucleon-nucleon inelastic cross section, which we can
take to be 42 mb at $\sqrt{s_{NN}}=200$ \cite{PDG98}.  The participant
number transverse density needed in Eqs.\ (\ref{nkk}) and (\ref{eq40})
along the jet trajectory can be similarly obtained from the Glauber
model to be
\begin{eqnarray}
\frac{dN_{\rm part}}{d{\bf b}'}({\bf b}') =AT({\bf b}_A') 
+B T({\bf b}_B'), 
\end{eqnarray}
where the transverse coordinates are given by $ {\bf b}_A'={\bf
b}'+{\bf b}/2, $ and $ {\bf b}_B'={\bf b}'-{\bf b}/2 $.  These
relations allows us to use Eq. (\ref{Pna}) to evaluate $P(N,\phi_s)$.
The distribution $P(N,\phi_s)$ can then be used in Eq.\ (\ref{nzeta})
to evaluate $R_{AA}(\phi_s)$, and in Eq.\ (\ref{eq37}) or (\ref{eq44})
to evaluate $\bar N_{k} (\phi_s)$.  After $\bar N_{k} (\phi_s)$ and
$R_{AA}(\phi_s)$ have been evaluated, we can average over all
azimuthal angles $\phi_s$ and we obtain the ridge particles [or
jet-(medium parton) collision] per trigger
\begin{eqnarray}
\langle N_{k} \rangle
=\int_0^{\pi/2} d\phi_s  \bar N_{k} (\phi_s)/(\pi/2),
\end{eqnarray}
and
\begin{eqnarray}
\langle R_{ AA} \rangle
=\int_0^{\pi/2} d\phi_s   R_{AA} (\phi_s)/(\pi/2),
\end{eqnarray}
which is usually expressed just as $R_{AA}$.

\begin{figure} [h]
\includegraphics[angle=0,scale=0.40]{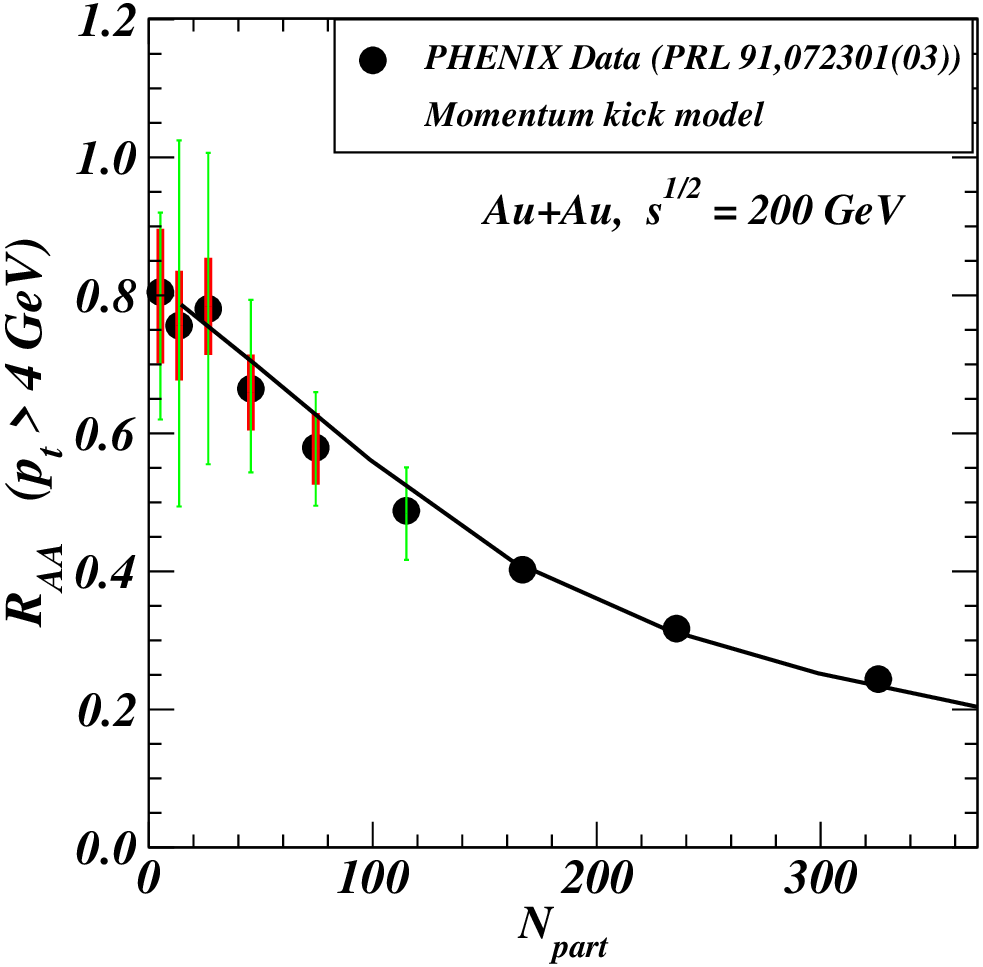}
\vspace*{0.0cm}
\caption{ (Color online) The ratio $R_{AA}$ of the high-$p_t$ $\pi^0$
trigger yield in $AA$ collision, as a function of the number of
participants $N_{\rm part}$. The solid curve is the theoretical result
from the momentum kick model, using $\zeta=0.20$ and $\sigma=0.7$ mb.
The data points are from PHENIX high-$p_t$ $\pi_0$ measurements for
AuAu collisions at $\sqrt{s_{NN}}=200$ GeV \cite{Adl03}.  }
\end{figure}

Eq.\ (\ref{eq11}) separates the ridge particle distribution into a
geometry-dependent part, $\langle f_R \rangle (2/3) {\langle N_{k}
\rangle}$, and the normalized ridge momentum distribution, $dF/d{\bf
p}$.  From the magnitude of the ridge yield, we have extracted
phenomenologically in Section VI the values of $\langle f_R \rangle
{\langle N_k \rangle}=3.8$ for central AuAu collisions at
$\sqrt{s_{NN}}=200$ GeV.  Ridge particles after production are
attenuated before reaching the detector.  It is reasonable to take the
average ridge particle attenuation factor $\langle f_R \rangle$ to be
the same as the average attenuation factor for jet component
particles, $f_J=0.632$, as both types of particles come out from the
interacting region.  We then get an estimate of $\langle N
\rangle=\langle N_k \rangle= 6.0$ as the total number of kicked partons
per trigger for the most-central AuAu collisions at 200 GeV.  For
numerical purposes, we shall use these average numbers as references,
keeping in mind however that they depend on the attenuation factor
$\langle f_R \rangle$ that may be uncertain.

\section{Comparison of Ridge Yield 
with Experimental Centrality Dependence}

For a given impact parameter and azimuthal angle $\phi_s$, the unknown
parameters are $\zeta$, and $\sigma$.  Although all quantities depend
on these two parameters, the quantity $R_{AA}$ for the quenching of
the trigger is more sensitive to $\zeta$, and the ridge yield per
trigger is more sensitive to $\sigma$.  We find that
the totality of experimental data of the centrality dependence of $R_{
  AA}$ and the centrality dependence of the ridge yield, can be
explained well when we use
\begin{eqnarray}
\label{zetapar}
\zeta=0.20, {\rm ~~and~~} \sigma =1.4 {\rm ~~mb} .
\end{eqnarray}

\begin{figure} [h]
\includegraphics[angle=0,scale=0.40]{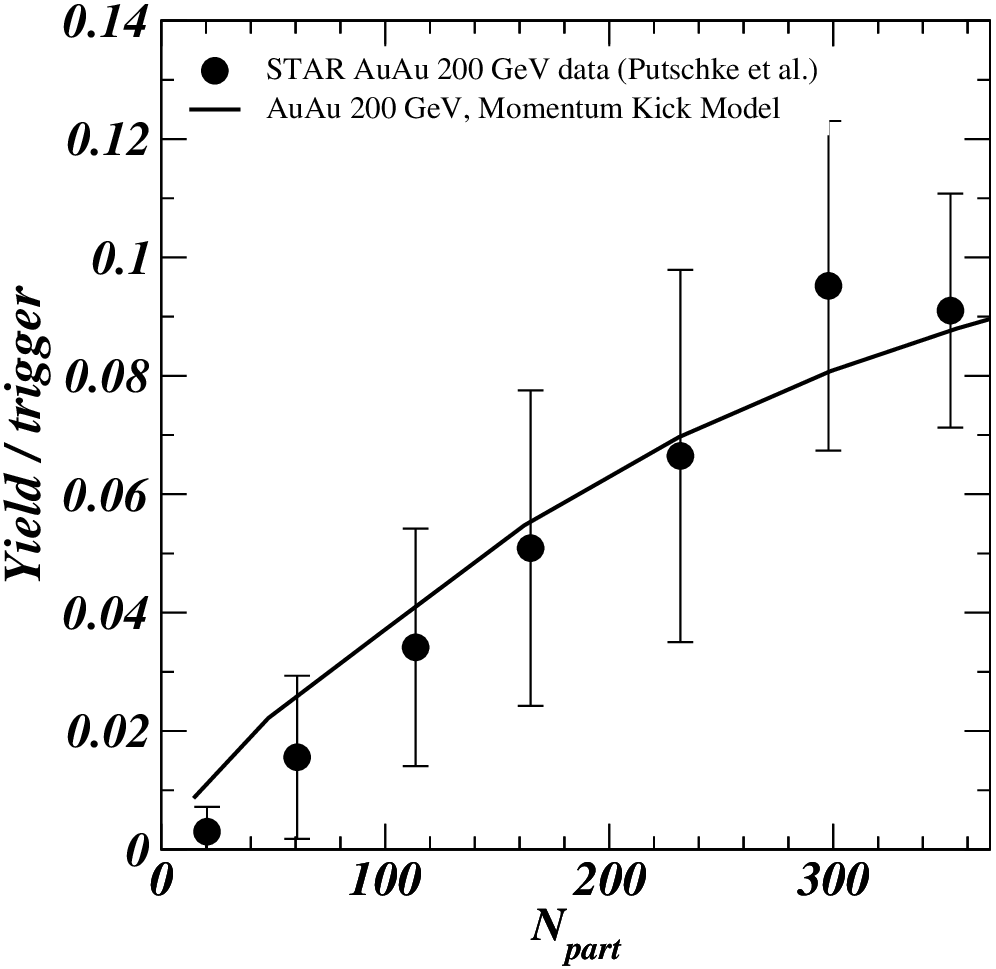}
\vspace*{0.0cm}
\caption{ (Color online) The ridge yield per trigger as a function of
the participant number $N_{\rm part}$ for nucleus-nucleus collisions
at $\sqrt{s_{NN}}$=200 GeV. The solid curve gives the theoretical
result for AuAu collisions in the momentum kick model.  The solid
circular data points are from the STAR Collaboration \cite{Put07}.  }
\end{figure}

We discuss here the comparison of theoretical results with
experimental data using the above two parameters for AuAu collisions
at $\sqrt{s_{NN}}$=200 GeV.  Solid circular points in Fig. 8 give
experimental PHENIX $R_{AA}$ data for high-$p_t$ $\pi^0$ yields
\cite{Adl03}.  Theoretical $R_{AA}$ result in the momentum kick model
obtained with Eq.\ (\ref{nzeta}) as a function of the participant
number is shown as the solid curve.  It gives good agreement with
experimental $R_{AA}$ data.  The quenching of the jet is well
accounted for in the momentum kick model.

The STAR AuAu ridge yield per trigger at $\sqrt{s_{NN}}$=200 GeV,
shown as solid circular points in Fig.\ 9, are taken from Fig.\ 2 of
\cite{Put07}.  They were obtained for $3 < p_{t,{\rm trig}} < 4$ GeV
and $2.0 < p_{t,{\rm associated}} < p_{t, {\rm trig}}$.  The solid
curve in Fig.\ 9 is the theoretical ridge yield per trigger for AuAu
collisions.  It has been normalized to match the data point (within
errors) for the most-central collision examined in Sections IV-VI.
Our comparison of momentum kick model results and the experimental
data in Fig.\ 9 indicates that theoretical ridge yields per
trigger agree with experiment.  It increases as the number of
participants increases.

The value of $\zeta=0.20$ is nearly the same as our earlier estimate
of $\zeta=0.22$ arising from collisional jet momentum loss alone.
This indicates that collisional momentum loss may contribute
the dominant component of the jet momentum loss, but more research on
theoretical predictions for $\zeta$ are needed to separate out the
different absorptive and radiative contributions.  The cross section
corresponds to a parton interacting radius of 0.21 fm, which means
that a parton having the entropy content of a hadron appears to the
jet probe as a strongly interacting scattering disk with a radius of
0.21 fm.

\section{Dependence of the Ridge Yield on Colliding Nuclei Masses and Energies}

Whereas our attention so far has been focused on AuAu collisions at
$\sqrt{_{NN}}=200$ GeV, we would like to investigate in this section
how the ridge yield scales with the mass numbers and the energies of
the colliding nuclei.  Experimental data for such an analysis have
been obtained by the STAR Collaboration with the acceptance region of
$3 < p_{t,{\rm trig}} < 4$ GeV and $2.0 < p_{t,{\rm associated}} <
p_{t, {\rm trig}}$ \cite{Bie07,Nat08}.  This region of acceptance is
slightly different from the acceptance region for Fig.\ 9 used in
Ref.\ \cite{Put07}.  The measurements of the ridge yield for AuAu and
CuCu collisions at $\sqrt{s_{NN}}=$200 and 62 GeV within the same
acceptance region in \cite{Bie07,Nat08} allows a consistent comparison
across mass numbers and energies of the colliding nuclei.  The
experimental ridge yield as a function of the participant numbers are
shown in Fig.\ 10 as solid points for AuAu collisions and open points
for CuCu collisions \cite{Bie07,Nat08}.  The circular data points are
for $\sqrt{s_{NN}}=200$ GeV and the square points are for 62 GeV.  One
notes that the ridge yield appears to increase with increasing number
of participants and increasing collision energies.  The ridge yield
for CuCu collisions is small and contains large systematic errors.

\begin{figure} [h]
\includegraphics[angle=0,scale=0.40]{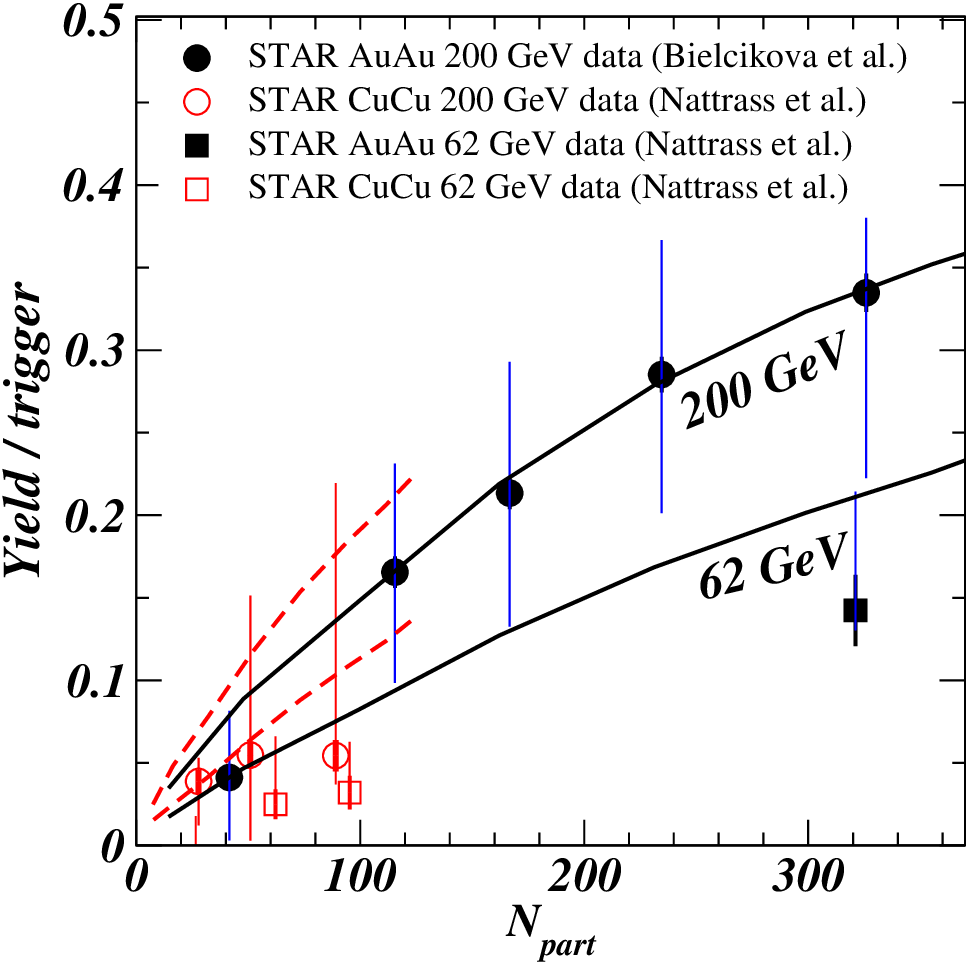}
\vspace*{0.0cm}
\caption{ (Color online) The ridge yield per trigger as a function of
the participant number $N_{\rm part}$ for nucleus-nucleus collisions
at $\sqrt{s_{NN}}$=200 and 62 GeV. The solid curves are theoretical
results for AuAu collisions and the dashed curves for CuCu collisions
in the momentum kick model.  The solid points represent AuAu data and
the open points represent CuCu data, from the STAR Collaboration
\cite{Bie07,Nat08}.  }
\end{figure}

We show in Fig. 10 the theoretical ridge yield for AuAu and CuCu
collisions as a function of the number of participants for
$\sqrt{s_{NN}}=200$ and 62 GeV.  The experimental and theoretical
ridge yields are matched for the most-central AuAu collision data
point at $\sqrt{s_{NN}}=$200 GeV.  Our comparison of momentum kick
model results and the experimental data at different energies,
different nuclear masses, and different participant numbers indicates
that the theoretical ridge yield agrees well with experiment.  For the
same nucleus-nucleus collision at different energies, the theoretical
ridge yield scales approximately with $\kappa$, the number of medium
partons produced per participant, which increases with increasing
collision energy as $(\ln \sqrt{s})^2$ \cite{Bus04}.  For the same
collision energy, the theoretical ridge yields per trigger for CuCu
collisions follow approximately those of the ridge yields for AuAu
collisions, when plotted in terms of the number of participants.

\section{Discussions and Conclusions}

The experimental near-side ridge data have guided us to the momentum
kick model as a description of the ridge phenomenon.  The narrow cone
of associated particles along the jet direction reveals that the
trigger particle is connected with the occurrence of a jet.  The yield
of the associated particles as increasing with increasing participant
numbers and the similarity of their inverse slope reveal that the
ridge particles come from medium particles.  The short-range behavior
of the strong interaction due to color screening and the narrow
azimuthal correlation of a ridge particle with the trigger particle
reveal further that the ridge particles and the jet are related by
collisions.  Hence, a picture of the momentum kick model emerges as a
plausible description of the ridge phenomenon.

In the momentum kick model, a jet parton produced in high-energy
heavy-ion collisions makes collisions with medium partons.  The kicked
medium partons subsequently materialize as ridge particles while the
jet loses energies and fragments into the trigger particle and other
fragmentation products.

The implementation of the momentum kick model can proceed numerically
by a Monte Carlo approach, following the trajectory of the jet and the
medium particles as the medium evolves.  The space-time dynamics of
the medium and the jet is a problem of great complexity and contains
many complex, unknown, and non-perturbative elements.  However, before
we implement such an elaborate undertaking, it is useful to explore
with simplifying assumptions whether the momentum kick model contains
promising degrees of freedom.

Following the dynamics of a jet and and its interaction with the
medium, we show how the ridge yield can be greatly simplified by using
the average distribution of the medium particles and the average
momentum kick.  We are then able to separate the ridge particle yield
into a factor which depends on the average number of partons kicked by
the jet and another factor related to the (average) momentum
distribution of the kicked parton after acquiring a momentum kick from
the jet.  The ridge particles therefore carries information on the
momentum distribution of the partons at the moment of jet-(medium
parton) collisions.  They also carry information on the (average)
magnitude of the momentum kick a medium parton acquires.  These
complications of space-time dynamics of medium and jet partons have
been subsumed under the probability distribution $P_N(N)$, which
depends on geometry, medium parton dynamics, jet parton trajectories,
and jet-(medium parton) cross sections.

The medium partons kicked by the jet materialize as ridge particles
can be used to extract the early parton momentum distribution.  The
extracted early parton momentum distribution provides valuable
information for the mechanism of early parton production and the later
evolution of the system toward the state of quark-gluon plasma.  For
central AuAu collision at $\sqrt{s_{NN}}=$ 200 GeV, we find the
extracted early parton momentum distribution to have a thermal-like
transverse distribution but a rapidity plateau structure whose width
decreases as the transverse momentum increases.  We should note that
plateau rapidity structure has been known in QCD particle production
experiments \cite{Aih88,Hof88,Pet88,Abe99,Abr99} and in QCD particle
production theories \cite{Cas74,Bjo83,And83,Won91,Won94}.  From this
viewpoint, the occurrence of a plateau structure at the early stage of
nucleus-nucleus collision should not come as a surprise.

The rapidity plateau distribution differs from the rapidity
distribution of the bulk matter which is found to have a Gaussian
shape \cite{Mur04,Ste05,Ste07}.  It is important to note that jets
occur at an early stage of the nucleus-nucleus collisions, whereas the
bulk medium properties are measured at the end-point of the
nucleus-nucleus collision.  A significant dynamical evolution must
have occurred between the early beginning of the nucleus-nucleus
collision and the end-point of the nucleus-nucleus collision.
Therefore the early parton momentum distribution near the beginning
stage of the nucleus-nucleus collision needs not be the same as the
bulk matter distribution at the end-point of the nucleus-nucleus
collision.  One expects that starting with a non-isotropic plateau
rapidity distribution that is much elongated in the longitudinal
direction, a collision of two partons with large and opposing
longitudinal momenta in adjacent spatial locations will redistribute
the partons from the longitudinal direction towards the transverse
directions, with a decrease in the longitudinal momenta of the
colliding partons.  Hence, the evolution will likely smooth out the
anisotropic plateau rapidity structure to a significant degree as time
proceeds.

The subject of our focus being the near-side ridge and jet quenching
in the early collision history, how the parton distribution function
$F({\bf r}, {\bf p}, t)$ evolves subsequently from the initial state
to the end-point of nucleus-nucleus collision is beyond the scope of
the present manuscript.  The complete problem of parton evolution is a
problem of great complexity
\cite{Mod08,Dum07,Dum08,Mro93,Reb04,Xu08,Mue00}, involving
perturbative and non-perturbative elements.  For example, in one of
the descriptions using the Color-Glass-Condensate treatment of the
initial conditions \cite{Dum08}, it is not well-understood even within
the Color-Glass-Condensate community how the initial large rapidity
correlations can evolve into to a thermal distribution in a short
period of time of 1-2 fm/c or to a Gaussian rapidity distribution at
the end-point of the nucleus-nucleus collision.  Some recent advances
suggest intrinsic color plasma instabilities that can lead to a
breaking of the boost invariance \cite{Dum07,Mro93,Reb04}, and other
investigations suggest the bottom-up scenario involving $gg\to ggg$
\cite{Xu08,Mue00}.  The extracted early momentum distribution of a
rapidity plateau obtained here serves to high-light the important and
unsolved issues of parton evolution that are left outstanding by the
present findings of this manuscript.

The momentum loss of the jet parton and the geometry of the jet
trajectory are other important aspects of the momentum kick model.
The magnitude of the momentum kick imparted onto the medium parton has
been found to be $q=1.0$ GeV per jet-(medium parton) collision.  This
momentum gain by the kicked parton is clearly related to the momentum
loss of the jet as a result of the jet-(medium parton) collisions.
One obtains a good phenomenological description of the experimental
data of the centrality dependence and collisional energy dependence of
$R_{ AA}$ and the ridge yield.  The extracted physical quantities
furnish important, albeit approximate, empirical data for future
investigations on the dynamics of parton production, parton evolution,
and jet energy loss.  The subject will come over and over again, each
time with more and more accuracy and refinement, as we go through our
course in physics.

The successes of the simplifying model indicates that the momentum
kick model contains promising degrees of freedom for the description
of the gross features of the ridge phenomenon and jet quenching.
There is however a limited range for the application of a completely
analytical formulation, as many refinements and improvements
necessitate additional degrees of freedom.  Among other things, we
envisage the need for a better description of the elementary
jet-(medium parton) collision process, a better description of the
dynamics of the medium, and the inclusion of effects of medium
transverse collective and elliptic flows that depend on the reaction
plane orientations and medium spatial locations.  There is the further
complication for intermediate $p_t$ trigger particles that some of the
trigger particles may arise not from the jets but from the medium
\cite{Jia08}.  A Monte Carlo implementation of the momentum kick model
that will allow the inclusion of many refinements and improvements,
and will therefore be of great interest.

\vspace*{0.3cm} The author wishes to thank Drs.\ Fuqiang Wang,
V. Cianciolo, Jiangyong Jia, Zhangbu Xu, and C. Nattrass for helpful
discussions and communications.  This research was supported in part
by the Division of Nuclear Physics, U.S. Department of Energy, under
Contract No.  DE-AC05-00OR22725, managed by UT-Battelle, LLC.


\begin{thebibliography}{99}
\bibitem{Not1} Note that the rapidity distributions along the
  sphericity axis and the thrust axis are nearly the same on almost
  all rapidity points except at $|y| \lesssim 0.1$ (see Figs. 40 and
  41 in Ref.\ \cite{Aih88}).  While the rapidity distribution obtained
  in the spericity axis system appears to be smooth around $y\sim 0$,
  the rapidity distribution obtained in the thrust axis system
  exhibits an abrupt decrease at $|y|\lesssim 0.1$.
\end{thebibliography}
\end{document}